\documentclass[apj]{emulateapj}
\newcommand{\mr}[1]{$\mathrm{#1}$}

\newcommand{\swift}{{\it Swift}}

\usepackage{float}
\usepackage{amsmath,amstext,amssymb}

\usepackage[usenames,dvipsnames,svgnames]{xcolor}
\usepackage[hyperfootnotes=false, linktocpage=true, colorlinks,linkcolor=Red, citecolor=blue]{hyperref}
\usepackage{graphicx}

\slugcomment{The Astrophysical Journal, Version of 6 May 2014}
\shorttitle{X-ray and H$\alpha$ Variability in NGC 1097}
\shortauthors{Schimoia et al.}
\begin{document}

\title{Short-Timescale monitoring of the X-ray, UV and broad double-peak emission line of the nucleus of NGC\,1097}


\author{Jaderson S.\ Schimoia, Thaisa Storchi-Bergmann}
\email{silva.schimoia@ufrgs.br}
\affil{Instituto de F\'isica, Universidade Federal do Rio Grande do Sul, Campus do Vale, Porto Alegre, RS, Brazil}
\author{Dirk Grupe\altaffilmark{1}}
\affil{Space Science Center, Morehead State University, 235 Martindale Dr.Morehead, KY 40351}
\author{Michael Eracleous\altaffilmark{2}$^{,}$\altaffilmark{3}}
\affil{Department of Astronomy and Astrophysics and Institute for Gravitation and the Cosmos, Pennsylvania State University, \\525 Davey Lab, University Park, PA 16802, USA}
\author{Bradley M. Peterson}
\affil{Department of Astronomy,  140 West 18th Avenue,  and the Center for Cosmology and AstroParticle Physics,  \\191 West Woodruff Avenue, The Ohio State University, Columbus, OH 43210, USA}
\author{Jack A. Baldwin}
\affil{Department of Physics and Astronomy, Michigan State University, East Lansing, MI 48864, USA}
\author{Rodrigo S. Nemmen\altaffilmark{4}$^{,}$\altaffilmark{5}$^{,}$\altaffilmark{6}}
\affil{Instituto de Astronomia, Geof\'{\i}sica e Ci\^encias Atmosf\'ericas, Universidade de S\~ao Paulo, S\~ao Paulo, SP 05508-090, Brazil}
\and
\author{Cl\'audia Winge}
\affil{Gemini South Observatory, c/o AURA Inc., Casilla 603, La Serena, Chile}
\altaffiltext{1}{Also, Swift Mission Operation Center, 2582 Gateway Dr., State College, PA 16801, USA}
\altaffiltext{2}{Also, Center for Relativistic Astrophysics, Georgia Institute of Tchnology, Atlanta, GA 30332, USA}
\altaffiltext{3}{Also, Department of Astronomy, University of Washington, Seattle, WA 98195, USA}
\altaffiltext{4}{Also, NASA Goddard Space Flight Center, Greenbelt, MD 20771, USA}
\altaffiltext{5}{Also, Center for Research and Exploration in Space Science \& Technology (CRESST)}
\altaffiltext{6}{Also, Department of Physics, University of Maryland, Baltimore County, 1000 Hilltop Circle, Baltimore, MD 21250, USA}
\begin{abstract}
Recent studies have suggested that the short-timescale ($\lesssim\,7$\,days) 
variability of the broad ($\sim$10,000\,km\,s$^{-1}$) double-peaked H$\alpha$ 
profile of the LINER nucleus of NGC\,1097 could be driven by a variable 
X-ray emission from a central radiatively inefficient accretion flow (RIAF). 
To test this scenario, we have monitored the NGC\,1097 nucleus in X-ray
and UV continuum with {\em Swift} and the H$\alpha$ flux and profile in 
the optical spectrum using SOAR and Gemini-South from 2012 August to 2013 February. 
During the monitoring campaign, the H$\alpha$ flux remained at
a very low level --- 3 times lower than the maximum flux observed in
previous campaigns and showing only limited ($\sim 20\%$) variability.  
The X-ray variations were small, only $\sim 13\%$ throughout the campaign, 
while the UV did not show significant variations. We concluded that the 
timescale of the H$\alpha$ profile variation is close to the sampling interval 
of the optical observations, which results in only marginal correlation between 
the X-ray and H$\alpha$ fluxes. 
We have caught the AGN in NGC\,1097 in a very low activity state, in which the ionizing source was very weak 
and capable of ionizing just the innermost part of the gas in the disk. Nonetheless, the data presented 
here still support the picture in which the gas that emits the broad double-peaked Balmer lines is illuminated/ionized 
by a source of high-energy photons which is located interior to the inner radius of the 
line-emitting part of the disk.
\end{abstract}


%
%
%
\section{Introduction}

The energy emitted by active galaxy nuclei (AGNs) is provided by
accretion of mass onto a supermassive black hole (SMBH), via an
accretion disk, whose emission is observed in the UV continuum of
Seyfert galaxies and quasars \citep[e.g.,][]{Frank}. The optical
spectra of some AGN show extremely broad
($\sim$10,000\,km\,s$^{-1}$) double-peaked emission lines that are thought
to originate in the outer extension of the accretion disk. Indeed, models
of ionized gas emission rotating in a relativistic Keplerian accretion
disk around a SMBH have been generally successful in accounting for the
double-peaked profiles \citep{Chen89, CeH89, SB03, Strateva03, Lewis10}.  These
models explain self-consistently most observable features of the
double-peaked profiles \citep{Eracleous03}.  Long time-scale monitoring
(years to decades) has revealed variability of the double-peaked
profiles in several objects, including the radio
galaxy 3C390.3 \citep{Shapovalova10}, the LINER NGC\,1097
\citep{SB03}, the radio galaxy Pictor\,A and other
radio-loud AGNs \citep{Gezari07, Lewis10}.

The variations in double-peaked profiles can provide
an effective probe of disk models and lead to understanding of
the physical mechanisms that cause profile variability. For this reason,
we have monitored the double-peaked H$\alpha$ profile 
in the nuclear spectrum of NGC\,1097.  We were
fortunate enough to obtain a few Gemini-South spectra separated by
less than a week, which revealed  much shorter timescale profile variability 
than had been seen before in this source.
Our observations constrain the variability timescale of the
integrated flux of the profile and the velocity separation between the
blue and red peaks to approximately 7 days \citep{Schimoia}. This is
the shortest timescale variability ever seen in the double-peaked
profile of this object, and coincides with the estimated light travel
time between the nucleus and the line-emitting region of the disk.

The short timescale variations suggest that the line emission is driven by the
central source. A number of studies have compared the available energy
due to local viscous dissipation in the accretion disk with the
observed luminosity of emission lines and concluded that there is
indeed the need of an external ionizing source to power the observed
luminosity of the double-peaked emission lines. This has been argued
by \citet{Eracleous94}, \citet{Strateva06,
Strateva08} and \citet{Luo}. In the case of NGC\,1097,  
\citet{Nemmen06} have proposed
that the ionizing source is a radiatively inefficient accretion flow
(RIAF) \citep{Narayan08} located in the center (or inner rim) of the
disk. The 7-day timescale is consistent with an origin of the
emission-line variability as a reverberation signal of the varying
X-ray emission from the inner RIAF in the line-emitting portion of the
disk.

In our previous studies of profile variability in the spectrum
of NGC\,1097 \citep{Schimoia} we also found
an inverse correlation between the flux of
the line and the velocity separation between the two peaks of the
profile, confirming the previous result reported by \citet{SB03}. This
inverse correlation also supports the reverberation scenario: when the
flux is higher, the RIAF at the center is more luminous and
illuminates/ionizes farther out in the accretion disk, where the
disk rotational velocities are
lower (thus the profile is narrower); conversely, when the flux from
the RIAF is lower, the disk emissivity is weighted more heavily
towards smaller radii where the velocities are higher, hence the
profile is broader. A similar beheavior is also observed in 3C\,390.3 by \citet[][see their Figure 4]{Shapovalova01}.

In order to test the hypothesis that the H$\alpha$ variability is
a reverberation response to variations of the X-ray and UV continuum
emitted by the inner RIAF, we undertook a campaign to monitor
the X-ray and H$\alpha$ emission in NGC 1097. From 2012 August to
2013 February, 
we monitored the emission from the nucleus of NGC\,1097 in three
different wavelengths bands.  The high-energy continuum (presumably
emitted by the RIAF) was monitored with the {\em Swift} satellite, using the
XRT telescope to obtain X-ray fluxes and the UVOT(M2) to obtain the UV
fluxes.  The spectral monitoring of the H$\alpha$ double-peaked
profile was performed using the GOODMAN spectrograph at the SOAR
telescope. In this contribution, we report on the results of this campaign.
In \S 2 we describe the
observations and the data reduction and in \S 3 we present the
measurements of the properties of the H$\alpha$ double-peaked profile
and the flux measurements of X-ray/UV bands as well as the main
results from our monitoring campaign. In \S 4, we discuss the results
and their physical implications to the reverberation scenario and the
emitting structure of the double-peaked profile.  Our conclusions are
summarized in \S 5.

%
%
%
\section{Observations and Data Reduction}

\subsection{X-Ray and Ultraviolet Observations with Swift}
Observations were made with X-ray and ultraviolet (UV) telescopes on {\it Swift}
between 2012 July 26 and 2013 January 30 (MJD\footnote{For brevity, we
use only the five least-significant digits of the modified Julian Date (MJD); MJD56134
refers to JD 2456134} 56134 to 56327). 

The \swift\ X-ray telescope \citep[XRT, ][]{burrows05} 
was operated in photon counting mode \citep{hill04}. 
The data were reduced by the task {\tt xrtpipeline} version 0.12.6., 
which is included in the HEASOFT package 6.12.
Source counts were measured inside a circle of radius $47''$ and 
the background was determined from 
a source-free region of radius of $235''$ using
the task {\tt xselect} (version 2.4b). 
Auxillary response files were created
using the XRT task {\tt xrtmkarf}. The spectra were rebinned with 20 counts
per bin using the task {\tt grppha} and the response files 
{\tt swxpc0to12s6\_20010101v013.rmf} were applied. 
The rebinned spectra were modelled over the range
 0.3--10~keV in XSPEC v.12.7 with a single power-law and 
correction for Galactic absorption 
corresponding to a hydrogen column density 
$N_{\rm H} = 2.03 \times10^{20}$\,cm$^{-2}$ \citep{kalberla05},
which is very  close to the value of
$N_{\rm H} = 2.3 \times 10^{20}$\,cm$^{-2}$
determined from a {\em Chandra} observation by \citet{Nemmen06}.

We analyzed all available data obtained with the \swift\ 
UV/Optical Telescope \citep[UVOT, ][]{roming05} during this period.
We restrict our attention to data obtained through the 
UVM2 (2246~\AA)  filter as it is the cleanest of the UVOT filters  \citep{breeveld10}.
We employed the UVOT software task 
 {\tt uvotsource} to extract counts within a circular region of 3$\farcs$0 
radius for the nucleus of NGC\,1097 and used $20''$ radius to determine the background.
 The UVOT count rates were aperture corrected and then converted into  
 magnitudes and fluxes  based on the most recent UVOT calibration as 
 described by 
\citet{poole08} and  \citet{breeveld10}. The flux measurements were corrected for
Galactic reddening ($E_{B-V}=0.027$\,mag) following 
the corrections given by \citet{roming09}, based on  standard reddening correction 
curves from \citet{cardelli89}. X-ray and UV measurements from
{\em Swift} are given in Table \ref{xray_uv_data}.

\begin{center}
\begin{deluxetable}{l c c c c}

\tablecolumns{5}
\tablecaption{X-Ray and UV Measurements from {\em Swift}}
\tablehead{
UT Date		&MJD 	&$F_{X}$ 		 &$\alpha_{X}$   	    &$F({\rm M2})$}
\startdata                                                                                         
2012 Jul 26 &56134.374   &$3.88^{+1.15}_{-0.55}$ &$0.58 \pm  0.32$ &$0.92 \pm 0.06$ \\
2012 Jul 30 &56138.928   &$3.47^{+0.60}_{-0.43}$ &$0.59 \pm  0.30$ &$1.05 \pm 0.07$ \\
2012 Aug 03 &56142.342   &$4.93^{+0.62}_{-0.58}$ &$0.83 \pm  0.40$ &$1.01 \pm 0.07$ \\
2012 Aug 07 &56146.819   &$5.94^{+0.82}_{-0.55}$ &$0.97 \pm  0.39$ &$1.01 \pm 0.07$ \\
2012 Aug 11 &56150.781   &$2.70^{+0.50}_{-0.30}$ &$1.15 \pm  0.46$ &$1.01 \pm 0.07$ \\
2012 Aug 15 &56154.631   &$5.17^{+0.66}_{-0.74}$ &$0.65 \pm  0.36$ &$1.00 \pm 0.06$ \\
2012 Aug 19 &56158.967   &$3.63^{+0.60}_{-0.31}$ &$0.72 \pm  0.33$ &$0.92 \pm 0.06$ \\
2012 Aug 23 &56162.712   &$4.20^{+0.77}_{-0.55}$ &$0.76 \pm  0.29$ &$0.88 \pm 0.06$ \\
2012 Aug 27 &56166.786   &$3.65^{+0.51}_{-0.29}$ &$0.91 \pm  0.30$ &$1.02 \pm 0.07$ \\
2012 Aug 31 &56170.586   &$4.00^{+0.61}_{-0.63}$ &$0.75 \pm  0.29$ &$0.99 \pm 0.06$ \\
2012 Sep 04 &56174.538   &$2.69^{+0.28}_{-0.23}$ &$1.17 \pm  0.32$ &$0.98 \pm 0.06$ \\
2012 Sep 08 &56178.213   &$2.95^{+0.40}_{-0.32}$ &$1.03 \pm  0.33$ &$0.98 \pm 0.06$ \\
2012 Sep 12 &56182.295   &$3.45^{+0.52}_{-0.43}$ &$0.80 \pm  0.34$ &$0.96 \pm 0.06$ \\
2012 Sep 16 &56186.561   &$3.80^{+0.76}_{-0.50}$ &$0.99 \pm  0.35$ &$0.98 \pm 0.06$ \\
2012 Sep 20 &56190.781   &$4.28^{+1.89}_{-0.56}$ &$1.12 \pm  0.90$ &$1.07 \pm 0.09$ \\
2012 Sep 24 &56194.715   &$8.41^{+4.98}_{-2.28}$ &$0.28 \pm  0.67$ &$0.98 \pm 0.09$ \\
2012 Oct 02 &56202.063   &$4.23^{+0.61}_{-0.49}$ &$0.71 \pm  0.28$ &$0.91 \pm 0.06$ \\
2012 Oct 06 &56206.406   &$3.32^{+0.71}_{-0.39}$ &$0.73 \pm  0.33$ &$0.96 \pm 0.06$ \\
2012 Oct 10 &56210.958   &$3.61^{+1.10}_{-0.46}$ &$0.76 \pm  0.40$ &$0.93 \pm 0.06$ \\
2012 Oct 14 &56214.900   &$3.24^{+0.34}_{-0.42}$ &$0.81 \pm  0.30$ &$0.94 \pm 0.06$ \\
2012 Oct 18 &56218.039   &$3.14^{+0.46}_{-0.52}$ &$0.69 \pm  0.32$ &$0.99 \pm 0.06$ \\
2012 Oct 22 &56222.231   &$4.35^{+0.46}_{-0.52}$ &$0.84 \pm  0.27$ &$0.95 \pm 0.06$ \\
2012 Oct 26 &56226.722   &$2.89^{+0.50}_{-0.40}$ &$0.91 \pm  0.36$ &$0.96 \pm 0.06$ \\ 
2012 Oct 30 &56230.864   &$2.52^{+0.46}_{-0.29}$ &$1.26 \pm  0.48$ &$0.89 \pm 0.06$ \\
2012 Nov 03 &56234.130   &$2.09^{+0.48}_{-0.26}$ &$1.37 \pm  0.46$ &$0.96 \pm 0.06$ \\
2012 Nov 07 &56238.066   &$3.39^{+0.57}_{-0.56}$ &$0.61 \pm  0.35$ &$0.97 \pm 0.06$ \\
2012 Nov 15 &56246.648   &$2.95^{+0.56}_{-0.44}$ &$0.87 \pm  0.38$ &$0.97 \pm 0.06$ \\
2012 Nov 19 &56250.828   &$3.15^{+0.39}_{-0.42}$ &$0.79 \pm  0.33$ &$0.95 \pm 0.06$ \\
2012 Nov 27 &56258.854   &$3.12^{+0.72}_{-0.52}$ &$0.78 \pm  0.36$ &$0.94 \pm 0.06$ \\ 
2012 Dec 01 &56262.323   &$2.87^{+0.47}_{-0.40}$ &$0.96 \pm  0.37$ &$0.85 \pm 0.05$ \\
2012 Dec 05 &56266.800   &$2.01^{+0.43}_{-0.25}$ &$0.90 \pm  0.36$ &$0.87 \pm 0.05$ \\ 
2012 Dec 09 &56270.465   &$3.05^{+0.51}_{-0.38}$ &$1.08 \pm  0.31$ &$0.95 \pm 0.06$ \\
2012 Dec 13 &56274.674   &$4.35^{+0.82}_{-0.60}$ &$0.77 \pm  0.39$ &$0.91 \pm 0.06$ \\
2012 Dec 17 &56278.610   &$3.00^{+0.68}_{-0.40}$ &$0.95 \pm  0.40$ &$0.88 \pm 0.06$ \\
2012 Dec 21 &56282.482   &$3.35^{+0.59}_{-0.40}$ &$1.04 \pm  0.37$ &$0.88 \pm 0.05$ \\
2012 Dec 25 &56286.829   &$5.05^{+1.08}_{-0.75}$ &$0.67 \pm  0.35$ &$0.93 \pm 0.06$ \\
2012 Dec 29 &56290.834   &$3.36^{+0.64}_{-0.66}$ &$0.76 \pm  0.32$ &$0.89 \pm 0.06$ \\
2013 Jan 02 &56294.041   &$2.78^{+0.62}_{-0.27}$ &$0.52 \pm  0.60$ &$0.88 \pm 0.06$ \\
2013 Jan 06 &56298.645   &$2.99^{+0.59}_{-0.49}$ &$0.62 \pm  0.32$ &$0.89 \pm 0.06$ \\
2013 Jan 10 &56302.618   &$2.04^{+0.46}_{-0.24}$ &$1.01 \pm  0.43$ &$0.87 \pm 0.06$ \\
2013 Jan 14 &56306.397   &$2.49^{+0.60}_{-0.38}$ &$0.69 \pm  0.43$ &$0.90 \pm 0.06$ \\
2013 Jan 18 &56310.286   &$2.90^{+0.46}_{-0.40}$ &$1.06 \pm  0.30$ &$0.97 \pm 0.06$ \\
2013 Jan 22 &56314.755   &$3.66^{+0.52}_{-0.35}$ &$1.00 \pm  0.28$ &$0.95 \pm 0.06$ \\
2013 Jan 26 &56318.159   &$2.62^{+0.37}_{-0.27}$ &$1.05 \pm  0.33$ &$0.93 \pm 0.06$ \\
2013 Jan 30 &56322.089   &$3.82^{+0.55}_{-0.32}$ &$0.74 \pm  0.29$ &$0.85 \pm 0.05$
\label{xray_uv_data}
\tablecomments{ Column (1) gives the date of observations while column (2) gives the Modified Julian Date (JD$-2400000.5$).
Column (3) gives the 0.3--10\,keV flux in 
units of $10^{-12}$\,erg\,s$^{-1}$\,cm$^{-2}$ and the X-ray
power-law slope appears in column (4). Column (5) gives the UV M2
($\sim2246\,$\AA) flux in units of $10^{-15}$\,erg\,s$^{-1}$\,cm$^{-2}$.}
\end{deluxetable}
\end{center}

\subsection{Optical Observations with SOAR}

The H$\alpha$ region of the optical spectrum was monitored at the SOAR Telescope
using the Goodman High-Throughput Spectrograph in long-slit mode. Observations were obtained
approximately every 7 days in queue mode from 
MJD56147 to MJD56319 for a total of 22 epochs (Program SO2012B-020). A few scheduled observations
were lost to poor weather so the actual mean interval between observations is $\sim 7.5$ days.
A log of observations appears
in Table \ref{obslog}.

The observations employed a 600\,l\,mm$^{-1}$ grating, yielding a
dispersion of 0.065\,nm\,pixel$^{-1}$ in the ``Mid" mode. 
A GG455 filter was used to block
second-order contamination. This setup provides wavelength coverage 450 -- 725\,nm 
and spectral resolution of $\sim 0.55$\,nm, measured as the FWHM of the lines in the arc spectrum.
The spectrograph slit was set to a projected width of $1\farcs03$ ($\sim80$\,pc at the galaxy), 
because it is usually larger
than the average seeing during the observations. The same slit
width was used in our previous observations (SB03), which 
allows us to scale the new spectra to match the 
the narrow emission line fluxes observed in our previous studies. 
This allows us to ``normalize" the spectra to a common flux scale.

\subsection{Gemini Data}
Motivated by the low H$\alpha$ flux in the SOAR spectra obtained early in this campaign
compared to what had been observed previously \citep{Schimoia},
a single long-slit spectrum was obtained
with the Gemini South Multi-Object Spectrograph (GMOS) on MJD56173. 
This observation was obtained as part of an active  ``Poor Weather'' project 
on the Gemini South telescope (Program GS-2012A-Q-86). 
For this observeration, we used a slit width of 1$\farcs$0, the B600 grating, and a GG455 filter.
The resulting spectrum covers the range from 510\,nm to 
800\,nm with a resolution of $\sim$\,0.45\,$\mathrm{nm}$. The data were reduced
using the standard procedures in the
\textit{IRAF}\footnote{IRAF is distributed by the National Optical Astronomy Observatories, 
which are operated by the Association of Universities for Research 
in Astronomy, Inc., under cooperative agreement with the National 
Science Foundation.} software package.
The resulting GMOS spectrum demonstrated
that despite the low H$\alpha$ flux, the 
double-peaked profiles obtained at SOAR were sufficiently reliable for 
detailed analysis.

\begin{deluxetable}{c c c c c c}
\centering
\tablecolumns{6}
\tablecaption{Observation Log}
\tablehead{
Telescope		&UT Date	&MJD		&Exposure		&P.A.\\
			&		&    		&Time (s)		&($^{\circ}$)}\\
\startdata
SOAR			& 2012 Aug 07	&56147.356	&3$\times$1200		&260\\
SOAR			& 2012 Aug 09	&56149.368	&3$\times$1200		&262\\
SOAR			& 2012 Aug 10	&56150.360	&3$\times$1200		&262\\
SOAR			& 2012 Aug 24	&56164.349	&2$\times$1200		&264\\
SOAR			& 2012 Aug 31	&56171.388	&2$\times$1200		&90\\
Gemini South		& 2012 Sep 03	&56173.305	&6$\times$600		&280\\
SOAR			& 2012 Sep 08	&56179.370	&3$\times$1200		&89\\
SOAR			& 2012 Sep 16	&56187.222	&3$\times$1200		&260\\
SOAR			& 2012 Sep 23	&56194.256	&2$\times$1500		&90\\
SOAR			& 2012 Sep 27	&56198.175	&5$\times$1200		&257\\
SOAR			& 2012 Oct 12	&56213.316	&3$\times$1200		&94\\
SOAR			& 2012 Oct 18	&56219.316	&3$\times$1200		&96\\
SOAR			& 2012 Oct 25	&56226.048	&3$\times$1200		&251\\
SOAR			& 2012 Oct 28	&56229.229	&3$\times$1200		&90\\
SOAR			& 2012 Nov 02	&56234.086	&3$\times$1200		&360\\
SOAR			& 2012 Nov 12	&56244.074	&3$\times$1200		&259\\
SOAR			& 2012 Nov 23	&56255.068	&3$\times$1800		&261\\
SOAR			& 2012 Dec 08	&56270.206	&3$\times$1800		&98\\
SOAR			& 2012 Dec 15	&56277.142	&3$\times$1800		&95\\
SOAR			& 2012 Dec 23	&56285.097	&3$\times$1200		&360\\
SOAR			& 2012 Dec 28	&56290.074	&3$\times$1200		&91\\
SOAR			& 2013 Jan 17	&56310.047	&4$\times$1200		&93\\
SOAR			& 2013 Jan 26	&56319.067	&3$\times$1200		&97
\label{obslog}
\tablecomments{The telescope used is given in column (1) and the date of observation
appears in column (2) while column (3) gives the Modified Julian Date (JD$-2400000.5$). Columns (4) and (5) give the exposure time  and slit position angle, respectively.}
\end{deluxetable}

%
%
%
\section{Data Analysis}
\label{results}
\subsection{The H$\alpha$ Profile}

\subsubsection{Nuclear extractions and the stellar population contribution}
\begin{figure*}[t!]
\centering
\includegraphics[width=7cm]{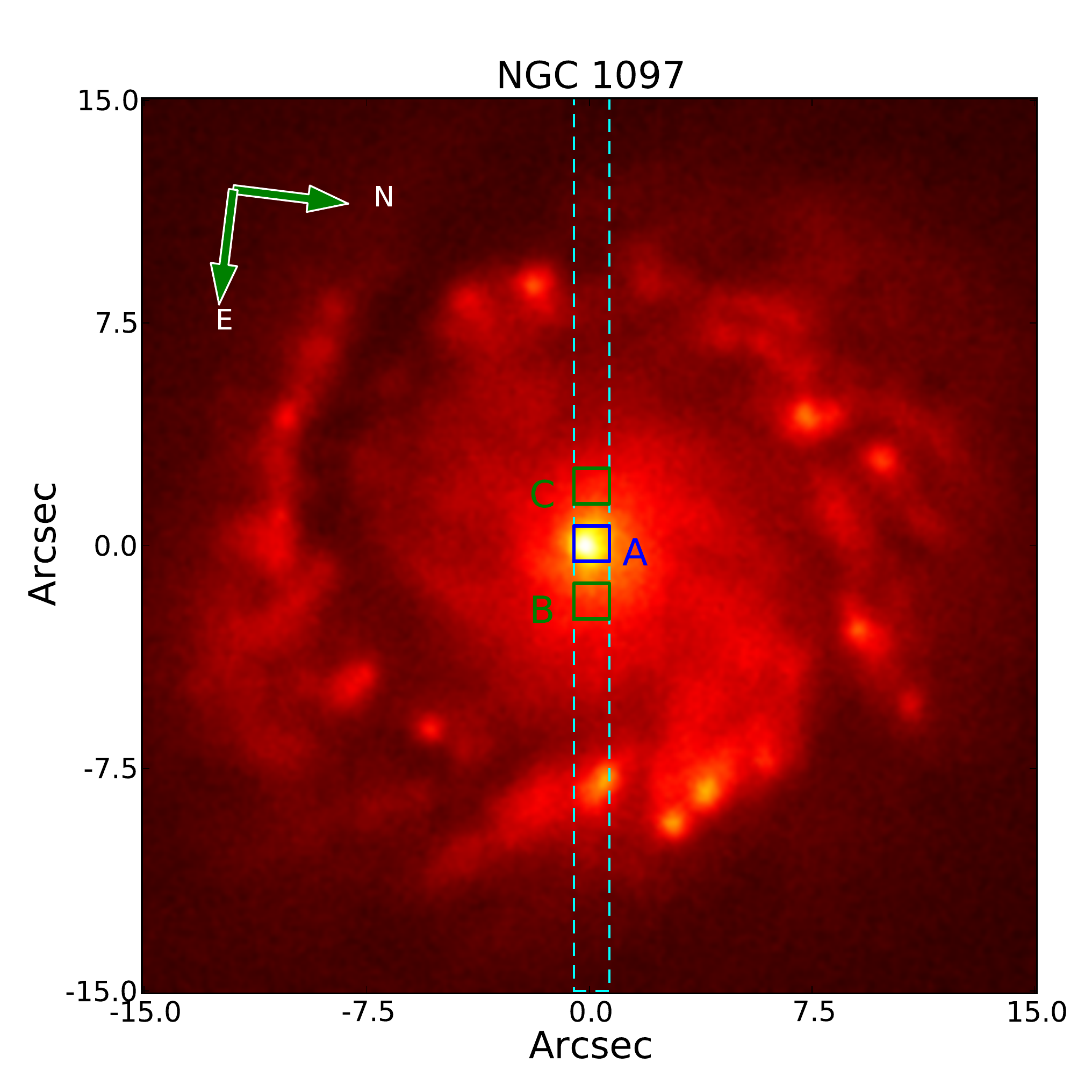}
\includegraphics[width=10.5cm]{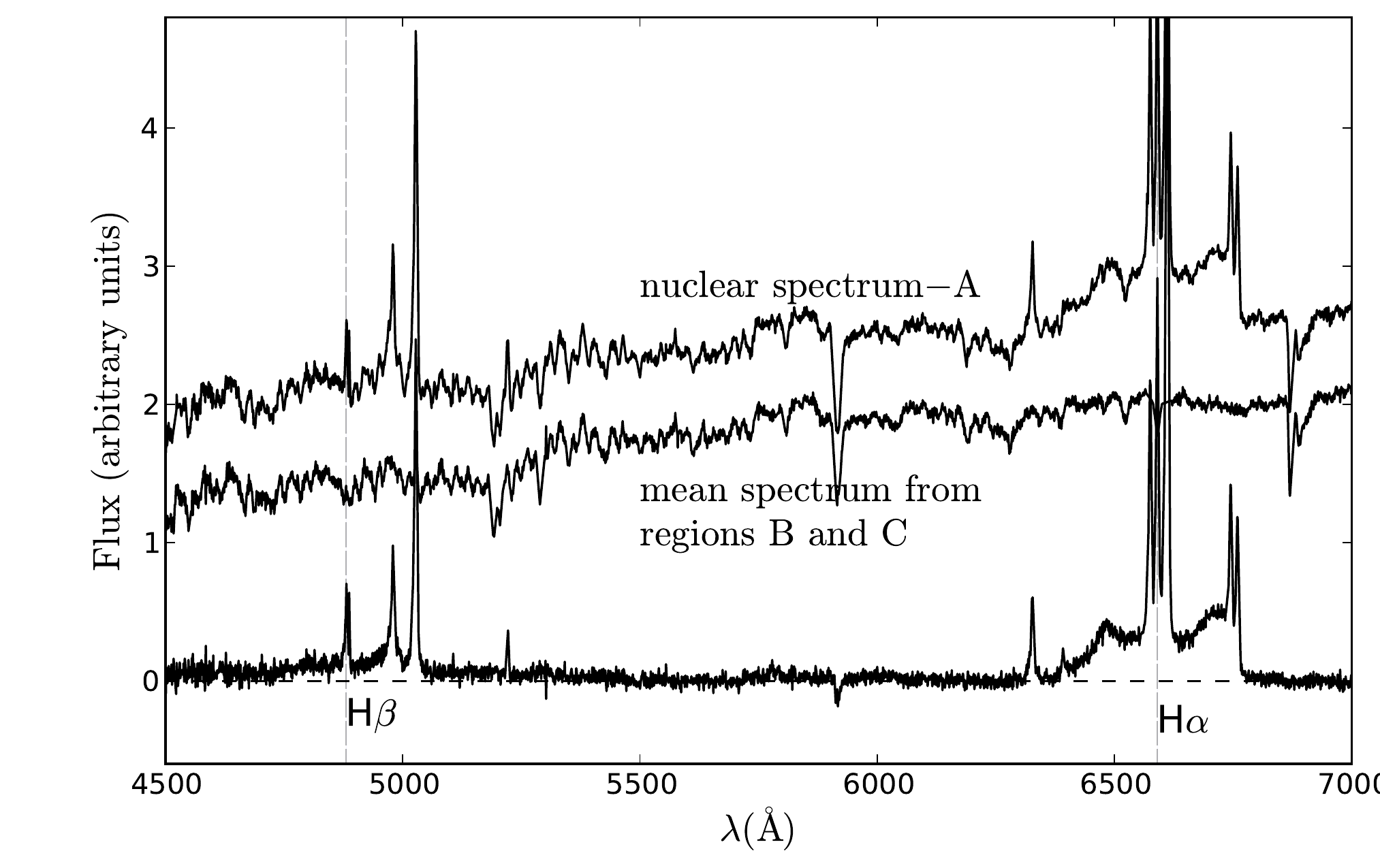}
\caption{
\textit{Left}: acquisition image of the nuclear region of  NGC\,1097 from the SOAR observation of MJD56194. The cyan dashed line represents
the slit (width of $1\farcs03$). The blue square (A) shows the extraction window ($1\farcs03 \times 1\farcs0$) of the nuclear spectrum. The green 
squares (B and C), centered at 2$\farcs$0 from the nucleus show the windows used to extract stellar population spectra. 
\textit{Right}: (a) top: nuclear spectrum extracted from window A; (b) middle: mean spectrum of B and C, adopted as representative of 
the nuclear stellar population; (c) bottom: difference between (a) and (b) (after scaling), which isolates the nuclear emission.
}
\label{acquisition}
\end{figure*}

We extracted the nuclear spectra using windows of $1\farcs03 \times1 \farcs0$ and $1\farcs0 \times1\farcs0$ for SOAR and Gemini data, respectively (Figure \ref{acquisition}). 
The nuclear extraction windows were centered at the peak of the continuum emission, which coincides with the location of the unresolved source of the broad H$\alpha$ line. 
Since the nuclear spectrum shows strong absorption lines from the underlying stellar population (Figure \ref{acquisition}),  we subtracted the stellar population contribution in order to isolate the H$\alpha$ profile.
For each epoch,  we extracted and then averaged two additional spectra
with extraction windows centered $2\farcs0$ away from the nuclear window (Figure \ref{acquisition}).  
The average extranuclear spectra display the same absorption features as the nuclear spectra,
and are only weakly ``contaminated" by narrow emission lines which were excised by using a synthetic spectrum obtained
by running the {\tt starlight-v04} code of \citet{Cid} as a template.
We assume that  this spectrum is representative of the nuclear stellar population and
scale and subtract it from the nuclear spectrum,
thus isolating the emission-line spectrum, as shown in Figure \ref{acquisition}.
As in our previous studies, the nuclear spectrum shows no detectable non-stellar continuum emission, only line emission.

We then normalized in flux the nuclear emission spectra using as reference a previous flux-calibrated
spectrum from 1991 November 2 \citep{SB93}, as we have in previous studies \citep{SB03,Schimoia}.
We assume that the fluxes in [N\,{\sc ii}]\,$\lambda\,\lambda$6548, 6584,  [S\,{\sc ii}]\,$\lambda
\lambda$6717, 6731, and the narrow component of H$\alpha$ 
have not varied since our 1991 observation because of the long light-travel time
across the region as well as the long recombination time at low densities (although
these assumptions may not hold for higher-ionization narrow lines,
see \citealt{Peterson13}).
Hereafter we refer to the flux-calibrated starlight-corrected nuclear emission spectra 
simply as the ``nuclear emission spectra''.

As we noted earlier, the broad H$\alpha$ flux is lower than it had been in our earlier observations, and consequently
some of the spectra are noisy. This leads to some ambiguity in measuring profile parameters (as described in the
next section) such as the wavelengths of the red and blue peaks. We therefore smoothed the
spectra with a Gaussian function. We experimented with Gaussian  smoothing functions between
one and five times the spectral resolution (i.e., FWHM between 5.5 and 27.5\,\AA), finally settling on
${\rm FWHM} = 11$\,\AA. This resulted in much better defined profile parameters for noisier spectra without
altering the measurements for the less noisy spectra.

\subsubsection{Salient features of the H$\alpha$ profile variations}
\begin{figure*}
\centering
\includegraphics[width=17cm]{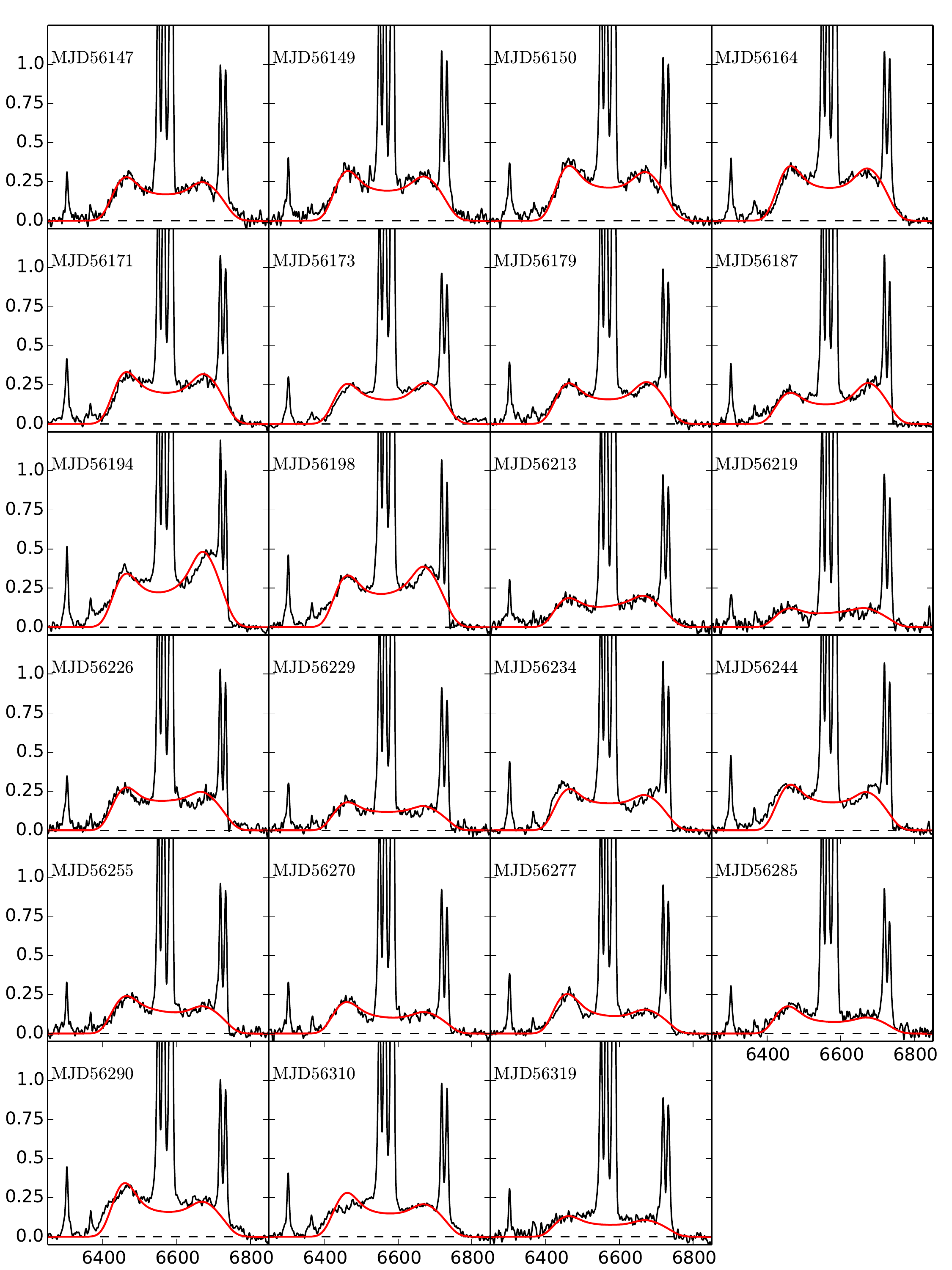}
\caption{
Resulting spectra of each epoch after the subtraction of the stellar population contribution and calibration through the fluxes of
the narrow emission lines (which do not vary on such short time scales). For each frame the vertical axis is flux in units of $10^{-15}$\,erg\,s$^{-1}$\,cm$^{-2}$\,\AA$^{-1}$ while
the horizontal axis is wavelength in units of \AA. The red solid line is the best fit of the accretion disk model to the data (see \S \ref{accretion_disk} for details).
}
\label{spectra}
\end{figure*}

The smoothed nuclear emission spectra are shown in Figure \ref{spectra}.
The most conspicuous characteristics of the evolution of the profile during the monitoring campaign are: 
\begin{enumerate}
\item  The double-peaked profile remained asymmetric, although the relative intensity of  the blue and red peaks changed during the campaign. At the beginning of the observations, MJD56147, 
the maximum flux of the blue peak, $F_B$, was slightly stronger than the
maximum flux of the red peak, $F_R$. The profile changed 
gradually until the red peak was much stronger than the blue peak by MJD56194. 
After that, the blue peak became increasingly prominent, becoming  stronger than 
the red peak by MJD56277.
This is similar to the results of our  previous studies that showed that the relative intensity of the blue and red peaks changes
on a timescale of months \citep{SB03, Schimoia}.

\item  Besides the changes in the relative intensity of the peaks, 
the integrated flux of the broad H$\alpha$ line, $F_{\rm broad}$, also varied significantly. 
The first remarkable 
rise in $F_{\rm broad}$ occurred in the interval MJD56187--56194,
when it changed from $62.5\ (\pm6.9) \times\,10^{-15}$\,erg\,s$^{-1}$\,cm$^{-2}$ to
$107.9\ (\pm4.4) \times\,10^{-15}$\,erg\,$s^{-1}$\,cm$^{-2}$, 
an increase of $\sim\,70\%$ in just six days. 
A second, even larger, flux increase of $\sim 90\%$ occurred 
in the interval MJD56285--56290, when the
H$\alpha$ flux increased from 
$43.8\ (\pm8.2) \times\,10^{-15}\,$erg\,s$^{-1}$\,cm$^{-2}$ to 
$83.4\ (\pm5.3) \times\,10^{-15}\,$erg\,s$^{-1}$\,cm$^{-2}$ in only five days.

\item After MJD56277, the red peak 
became less well-defined, with a shape more like a ``plateau''. 
Similar behavior is seen on the blue side of the profile in the observation of MJD56310.
By the final observation of the campaign on MJD56319, the double-peaked nature of the 
profile is barely discernible.
\end{enumerate}

\

\subsubsection{Model-independent parameterization of the douple-peaked profile}
\label{proprieties}

\begin{deluxetable*}{l c c c c c c c c}
\tablecolumns{8}
\tablecaption{Measurements of the characteristics of the double-peaked profile}
\tablehead{
UT Date & MJD & $F_{\rm broad}$	&$\mathrm{\lambda_{i}}$ &$\mathrm{\lambda_{f}}$ & $\mathrm{\lambda_{B}}$ & $\mathrm{\lambda_{R}}$ & $F_{B}$ & $F_{R}$}\\
\startdata
2012 Aug 07	&56147.356 &$71.9  \pm 8.9 $ &6381 &6784 &$6469.6 \pm  1.1$ &$6665.3   \pm  1.8$  &$0.264  \pm 0.088$ &$0.231 \pm 0.087$ \\ 
2012 Aug 09	&56149.368 &$87.6  \pm 10.5$ &6371 &6793 &$6467.0 \pm  4.1$ &$6666.6   \pm  5.3$  &$0.304  \pm 0.100$ &$0.279 \pm 0.099$ \\ 
2012 Aug 10	&56150.360 &$93.3  \pm 8.0 $ &6361 &6793 &$6469.9 \pm  2.7$ &$6661.2   \pm  4.3$  &$0.350  \pm 0.074$ &$0.295 \pm 0.073$ \\ 
2012 Aug 24	&56164.349 &$90.3  \pm 5.7 $ &6367 &6781 &$6473.5 \pm  3.4$ &$6660.0   \pm  5.5$  &$0.309  \pm 0.054$ &$0.295 \pm 0.054$ \\ 
2012 Aug 31	&56171.388 &$85.4  \pm 5.4 $ &6364 &6789 &$6476.8 \pm  2.3$ &$6672.4   \pm  1.3$  &$0.291  \pm 0.050$ &$0.275 \pm 0.050$ \\ 
2012 Sep 03	&56173.305 &$70.9  \pm 7.8 $ &6365 &6772 &$6466.8 \pm  1.2$ &$6685.3   \pm  7.7$  &$0.220  \pm 0.078$ &$0.236 \pm 0.074$ \\ 
2012 Sep 08	&56179.370 &$70.4  \pm 7.1 $ &6365 &6772 &$6463.9 \pm  1.2$ &$6684.3   \pm  3.1$  &$0.238  \pm 0.070$ &$0.247 \pm 0.070$ \\ 
2012 Sep 16	&56187.222 &$62.5  \pm 6.9 $ &6370 &6755 &$6464.8 \pm  2.2$ &$6683.2   \pm  1.9$  &$0.197  \pm 0.070$ &$0.248 \pm 0.071$ \\ 
2012 Sep 23	&56194.256 &$107.9 \pm 4.4 $ &6356 &6775 &$6467.4 \pm  2.1$ &$6679.3   \pm  2.0$  &$0.337  \pm 0.042$ &$0.438 \pm 0.042$ \\ 
2012 Sep 27	&56198.175 &$92.4  \pm 7.1 $ &6365 &6776 &$6465.1 \pm  4.6$ &$6669.3   \pm  3.7$  &$0.301  \pm 0.068$ &$0.354 \pm 0.068$ \\ 
2012 Oct 12	&56213.316 &$51.2  \pm 9.5 $ &6375 &6783 &$6460.5 \pm  2.1$ &$6685.4   \pm  2.5$  &$0.169  \pm 0.092$ &$0.182 \pm 0.093$ \\ 
2012 Oct 18	&56219.316 &$30.6  \pm 9.0 $ &6386 &6763 &$6457.8 \pm  1.8$ &$6683.7   \pm  2.4$  &$0.118  \pm 0.045$ &$0.101 \pm 0.095$ \\ 
2012 Oct 25	&56226.048 &$66.8  \pm 8.3 $ &6374 &6800 &$6465.2 \pm  2.3$ &$6682.9   \pm  2.4$  &$0.246  \pm 0.078$ &$0.206 \pm 0.078$ \\ 
2012 Oct 28	&56229.229 &$42.5  \pm 8.4 $ &6375 &6786 &$6453.4 \pm  2.5$ &$6687.3   \pm  4.6$  &$0.171  \pm 0.081$ &$0.134 \pm 0.081$ \\ 
2012 Nov 02	&56234.086 &$66.5  \pm 4.5 $ &6372 &6780 &$6458.3 \pm  3.6$ &$6685.2   \pm  2.7$  &$0.269  \pm 0.044$ &$0.222 \pm 0.044$ \\ 
2012 Nov 12	&56244.074 &$75.5  \pm 6.4 $ &6363 &6785 &$6475.2 \pm  1.6$ &$6678.2   \pm  3.9$  &$0.270  \pm 0.060$ &$0.256 \pm 0.061$ \\ 
2012 Nov 23	&56255.068 &$53.6  \pm 8.1 $ &6377 &6751 &$6463.9 \pm  3.5$ &$6677.2   \pm  5.1$  &$0.219  \pm 0.086$ &$0.169 \pm 0.086$ \\ 
2012 Dec 08	&56270.206 &$42.7  \pm 6.7 $ &6375 &6763 &$6464.0 \pm  0.5$ &$6669.0   \pm  2.1$  &$0.198  \pm 0.071$ &$0.124 \pm 0.071$ \\ 
2012 Dec 15	&56277.142 &$47.2  \pm 4.8 $ &6369 &6762 &$6471.1 \pm  2.8$ &$6670.2   \pm  2.0$  &$0.226  \pm 0.049$ &$0.133 \pm 0.049$ \\
2012 Dec 23	&56285.097 &$43.8  \pm 8.2 $ &6381 &6787 &$6468.9 \pm  2.9$ &$6672.0   \pm  2.8$  &$0.179  \pm 0.090$ &$0.123 \pm 0.046$ \\ 
2012 Dec 28	&56290.074 &$83.4  \pm 5.3 $ &6363 &6794 &$6467.6 \pm  1.8$ &$\ldots$             &$0.307 \pm 0.1$35&$\ldots$\\                            
2013 Jan 17	&56310.047 &$67.4  \pm 5.7 $ &6367 &6788 & $\ldots$         &$6672.2 \pm 5.1$&$\ldots$&$0.202 \pm 0.047$\\ 
2013 Jan 26	&56319.067 &$40.5  \pm 8.0 $ &6376 &6753 &$6496.32 \pm 5.2$ &$6686.7 \pm 8.5$ &$0.140 \pm 0.047$&$0.125 \pm 0.106$
\enddata
\tablecomments{Column (1) gives the date of observations while column (2) gives the Modified Julian Date (JD$-2400000.5$).
In column (3) is shown the integrated flux of the double-peaked profile in units of
$\mathrm{10^{-15}\ erg\,s^{-1}\,cm^{-2}}$.
In column (4) is given the initial wavelength of the double peaked profile and in (5)
is the corresponding final wavelength, both are given in \AA.
In columns (6) and (7) are given, respectively, the wavelengths in \AA\ where
the maximum intensity of the blue and the red peaks occur.
In columns (8) and (9) are given the values of the maximum intensity
of the blue and red peaks, respectively, in units of $\mathrm{10^{-15}\ erg\,s^{-1}\,cm^{-2}\,{\AA}^{-1}}$.
}
\label{tabledata}
\end{deluxetable*}

\begin{figure}
\centering
\includegraphics[width=8.5cm]{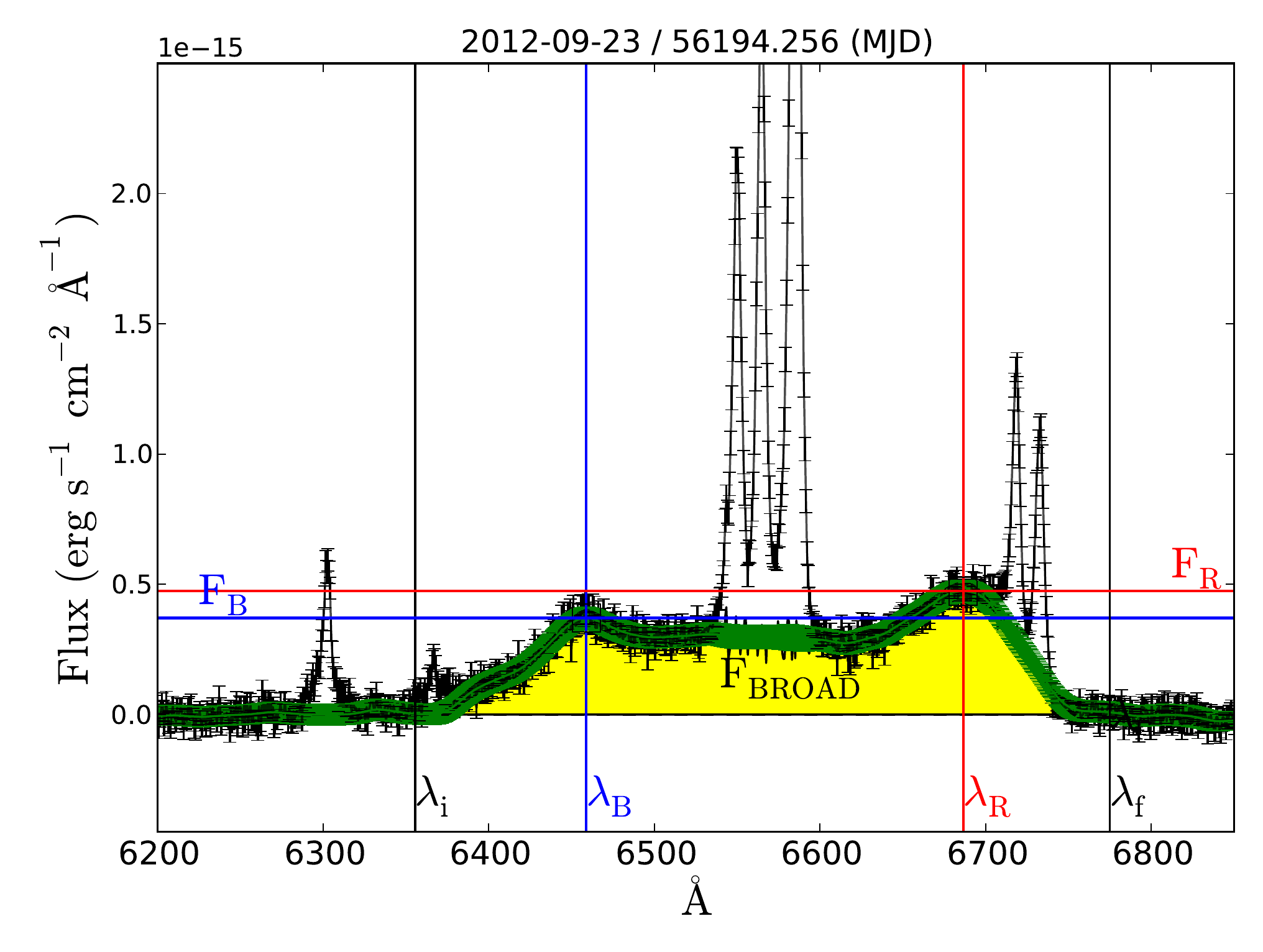}
\caption{Visual representation of the measured properties of the H$\alpha$ double-peaked profile for the 
observation of MJD56194. The vertical black lines mark $\lambda_i$ and $\lambda_f$ (respectively the lower and upper wavelength limits
of the double-peaked profile). The horizontal blue line marks the maximum flux value for the blue peak, $F_B$, while the vertical blue line marks $\lambda_B$, the wavelength where $F_B$ occurs. The maximum flux of the red peak $F_R$ is marked by the horizontal red line, while $\lambda_R$ is marked by the vertical red line. The green strip represents
the gaussian smoothing of the spectrum. The yellow region represents $F_{\rm broad}$, after subtracting the contribution of the narrow emission lines.}
\label{measure}
\end{figure}

Prior to making measurements to parameterize the broad H$\alpha$ profile, we must first
remove the very strong narrow emission lines that are still present in the spectrum. We 
attempted to remove these features by simultaneously fitting three Gaussians to
the H$\alpha$ narrow component and the [N\,{\sc ii}]\,$\lambda\,\lambda$6548, 6584 lines
and two more Gaussians to the [S\,{\sc ii}]\,$\lambda\,\lambda$6717, 6731 doublet.
After removing the narrow components, we made the following measurements:
$\lambda_{i}$ is the initial wavelength, defined as
the wavelength at which the flux of the blue side reaches zero intensity within the uncertainties. Similarly, $\lambda_f$ is the 
final wavelength where the flux of the red side of the profile reaches zero intensity. 
Together these two parameters 
define the wavelength range of the double-peaked profile. $F_B$ is the maximum intensity of the blue peak (or blue side) of the profile and $\lambda_{B}$ is the wavelength where this maximum occurs, while $F_R$ is the maximum intensity of the red peak (or red side) of the profile and $\lambda_{R}$ the wavelength where it occurs. We obtained the integrated flux in
the broad component of H$\alpha$ $F_{\rm broad}$ by integrating the flux underneath the profile 
from $\lambda_i$ to $\lambda_f$.  Figure \ref{measure} illustrates all these properties, 
including the results of fitting and subtracting the contribution of the narrow lines.
The resulting measurements are listed in Table \ref{tabledata}.
For most spectra, it was possible to make reliable measurements of the properties describe above. However, the irregular profiles in the final observations  (from MJD56285 on)
preclude characterization of the blue and red peaks.

\subsection{Time-Series Analysis}

\begin{figure*}
\centering
\includegraphics[width=17.3cm]{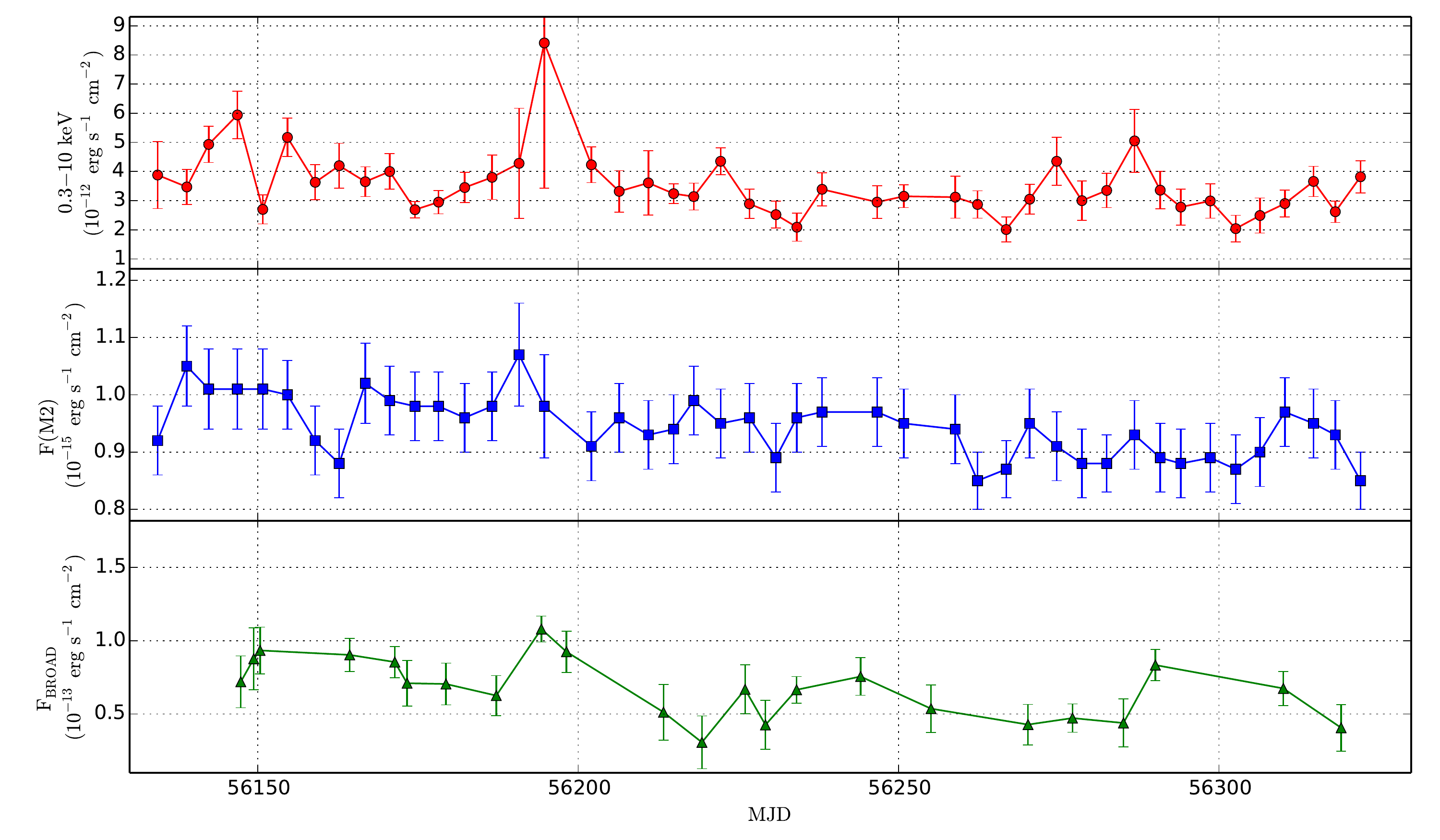}
\caption{
Light curves. \textit{Top}: integrated flux of the {\em Swift}/XRT X-ray
band, 0.3 -- 10 keV. \textit{Middle}: UV integrated flux, from the M2 band ($\sim 2246$\,\AA)
of {\em Swift}/UVOT. \textit{Bottom}:
integrated H$\alpha$ broad-line  flux. 
The largest average variation was observed for the H$\alpha$ profile, of ~20\%, 
followed by the X-rays, which varied by ~13\%, while the UV did not show significant variation.
}
\label{light_curves}
\end{figure*}

In this section, we characterize the variability of the X-ray and UV continuum and
the integrated flux $F_{\rm broad}$ of the broad H$\alpha$ profile.
Figure \ref{light_curves} shows the corresponding light curves obtained over the complete monitoring campaign.

The mean cadence of the observations during the monitoring campaign was based on the results from \citet{Schimoia}.
In this previous work, we did not observe any significant H$\alpha$ variations  within a time interval of seven days.
We therefore planned the 
monitoring campaign to obtain an optical spectrum with SOAR once every 8 days.
As the X-rays usually show shorter variability timescale  than the optical in AGNs,
we requested a {\it Swift} observation once every four days whenever  possible.

For the high-energy continuum emission, 45 {\it Swift} observations were obtained in each of two bands,
the XRT ($0.3$--$10$\,keV) and UVOT/M2 band ($\lambda_0 \approx 2246$\,\AA, ${\rm FWHM} \approx 510$\,\AA), 
with a mean time interval of $\sim 4.2$ days and minimum and maximum time intervals of $\sim 3$ and 8 days, respectively.
The $H\alpha$ light curve comprises
23 observations with a mean time interval between them of $\sim 7.5$ days, with minimum and maximum time intervals between
two consecutive observations of $\sim 1$ and 20 days, respectively.

We characterize the variability of each light curve using the same parameters as \cite{OBrien98}.
The results are given in Table \ref{var_data}, where $N$ is the number of epochs,  
$\overline{F}$ and $\sigma_{F}$
are, respectively the mean flux and standard deviation, and 
$\Delta$ is the mean measurement uncertainty. The parameter $F_{\rm var}$ is an estimate of the amplitude
of the variability relative to the mean flux, corrected by the uncertainties in the measurements, i.e.,
\begin{equation}
 F_{\rm var}=\frac{\sqrt{ {\sigma_{F}}^{2} - \Delta^{2}} }{\overline{F}}.
\end{equation}
The parameter 
$R_{\rm max}$ is the ratio between the maximum and minimum fluxes in the light curve.

\begin{deluxetable}{l c c c c c c}[ht]
\centering
\tablecolumns{8}
\tablecaption{Variability Parameters}
\tablehead{
Feature                                 &$N$      & $\overline{F}$    &$\sigma_{F}$   &$\Delta$   &$F_{\rm var}\,\mathrm{(\%)}$  &$R_{\rm max}$} \\
\startdata                                                                                                                        
XRT\tablenotemark{a}                    &45     &3.54               &1.1            &1.0        &13.18                     &4.18       \\
F(M2)\tablenotemark{b}                  &45     &9.44               &0.52           &0.62       &INDEF                     &1.26       \\
$F\mathrm{_{broad}}$\tablenotemark{b}   &23     &6.72               &2.01           &1.43       &20.44                     &3.52
\enddata
\tablenotetext{a}{$\overline{F}$, $\sigma_{F}$, and $\Delta$  are in units of $10^{-12}\,\mathrm{erg\,s^{-1}\,cm^{-2}}$.}
\tablenotetext{b}{$\overline{F}$, $\sigma_{F}$, and $\Delta$  are in units of $10^{-14}\,\mathrm{erg\,s^{-1}\,cm^{-2}}$.}
\label{var_data}
\end{deluxetable}

\subsubsection{X-Ray}

The X-ray variations  (Figure \ref{light_curves}) 
were of very low amplitude, with $F_{var} \approx 13\%$.
There were two periods of large changes, a large decrease followed by a large increase in flux around MJD56150
and another increase followed by a rapid decrease around MJD56194. 
These variations imply that the X-ray emitting structure must be smaller than four light days.
We note that there is an ultra-luminous X-ray source (ULX) located approximately
27$^{\prime\prime}$ away from the nucleus, as pointed out by \citet{Nemmen06}. As the {\em Swift} measurements employ an aperture of 47$^{\prime\prime}$ radius, this source is included in our measured X-ray flux. We have used the information provided by \citet{Nemmen06} to estimate an upper limit of $\sim8\%$ to the flux contributed by this ULX to our measurements.
If we correct our measurements by this constant contribution, we expect that
$F_{var} \approx 15\%$.

\subsubsection{UV}
The $F_{var}$ parameter for the M2 Swift band is
undefined because the variations in the UV flux are smaller than the 
uncertainties of these measurements.

\citet{Nemmen06} have used an {\em HST} STIS UV/optical spectrum of
the nucleus of NGC\,1097 as part of the nuclear spectral energy distribution (SED). With
data from radio wavelengths to X-rays, they then modeled
the nuclear SED as originating in a RIAF
plus the contribution from a radio jet, including also the continuum
of the part of the disk which emits also the double-peaked line.
Besides these components, it is also necessary to include the
contribution of a nuclear starburst \citep{SB05} to reproduce the UV
spectral range. The derived contributions of each of these structures
to the luminosity at 2246\,\AA\ are
$L_{\rm RIAF}=1.26\times10^{40}$\,erg\,s$^{-1}$ from the RIAF, 
$L_{\rm jet}=2.57\times10^{39}$\,erg\,s$^{-1}$ from the radio jet, 
$L_{\rm disk}=5.25\times10^{34}$\,erg\,s$^{-1}$ from the thin disk, and
$L_{\rm star}=2.04\times10^{40}$\,erg\,s$^{-1}$
from the compact nuclear starburst. We thus conclude that, in the observed UV band, the nuclear starburst, which has a projected distance to the nucleus smaller than 9 pc \citep{SB05},
dominates the luminosity. As the light from the starburst does not vary on the timescales we are considering, it can be concluded that any eventual 
variation of the AGN in this band is probably heavily diluted by the starburst continuum and cannot be detected. 

\subsubsection{The H$\alpha$ Emission Line}

As we can see in Figure \ref{spectra}, the H$\alpha$ profile and flux have varied
on the shortest timescales probed by the observations. The broad-component flux
$F_{\rm broad}$ varied overall by approximately $\sim20$\% ($F_{var}=20.44$\%) with respect to the mean.
The two most remarkable increases in $F_{\rm broad}$ occurred in the interval
MJD56187--56194 and  MJD56285--56290, when the line flux increased by 
$\sim 70\,\%$ in 7 days and by $\sim\,90\,\%$ in 5 days, respectively, as seen in 
Figures \ref{spectra} and \ref{light_curves}.
We conclude that a substantial portion of 
the broad-line luminosity comes from a region that has a size less than 5 light days.


\section{DISCUSSION}

\label{delV}
The double-peaked H$\alpha$ profile of NGC\,1097 has been monitored by our group 
for more than 20 years \citep{SB93, SB95, SB97, SB03}. Recently we reported the discovery of short timescale 
variability \citep{Schimoia} of $\sim 7$ days,  consistent with the 
shortest timescale variations detected here of  $\lesssim 5$ days.
Another previous finding  was an inverse correlation between F$_{\rm broad}$ and the velocity
separation between the blue and red peaks, $\Delta {V}= V_B - V_R$.
This inverse correlation supports the reverberation scenario:
when the ionizing flux is low, the innermost parts of the accretion disk are
relatively more important and the profile becomes weaker and broader, when the
ionizing flux is high the inner disk becomes more highly ionized, therefore the bulk of the Balmer line emission moves to larger radii \citep{Dumont} and the profile 
becomes both stronger and  narrower.  

\begin{figure*}
\centering
\includegraphics[width=7.1cm]{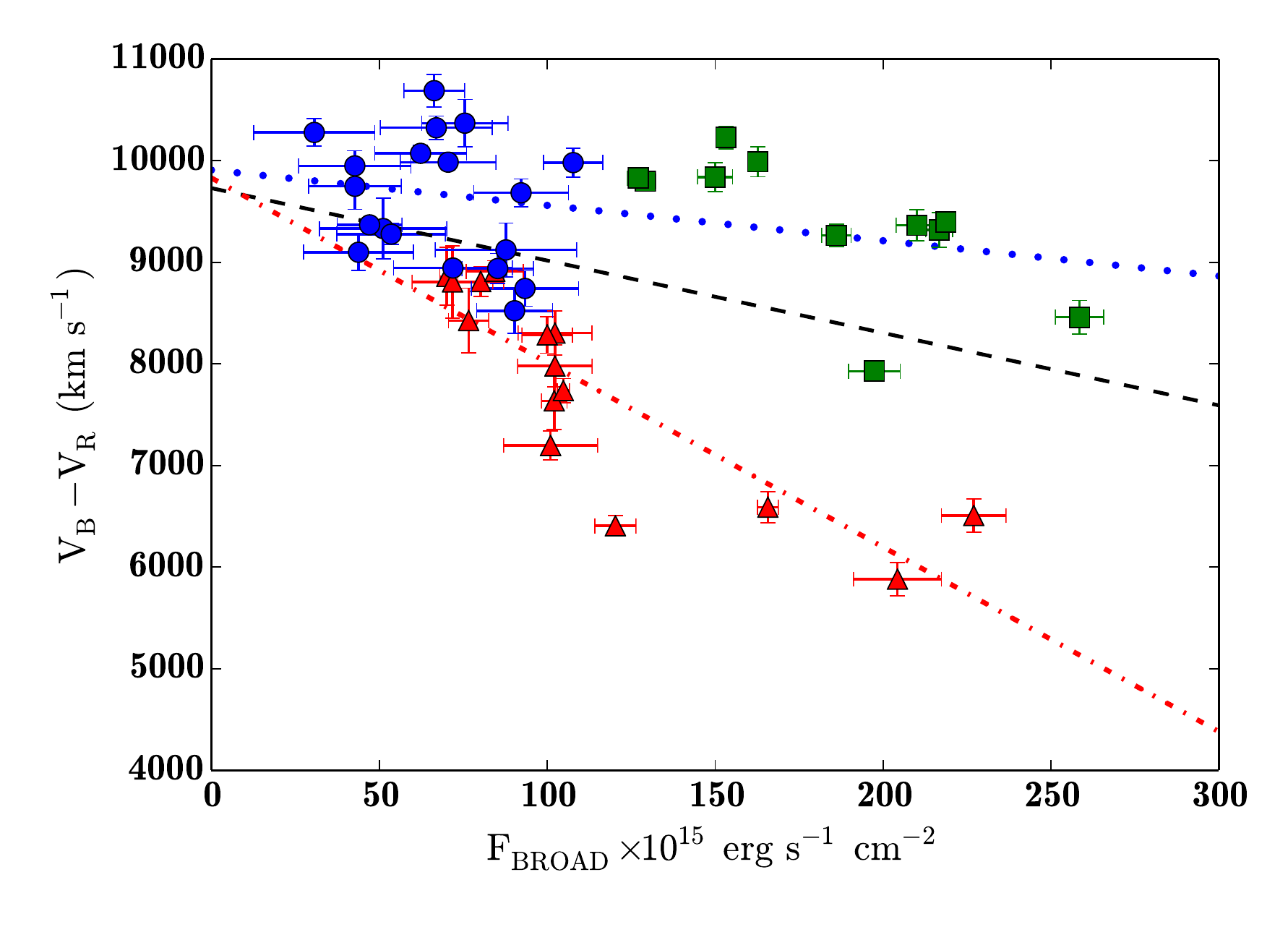}
\includegraphics[width=10.cm]{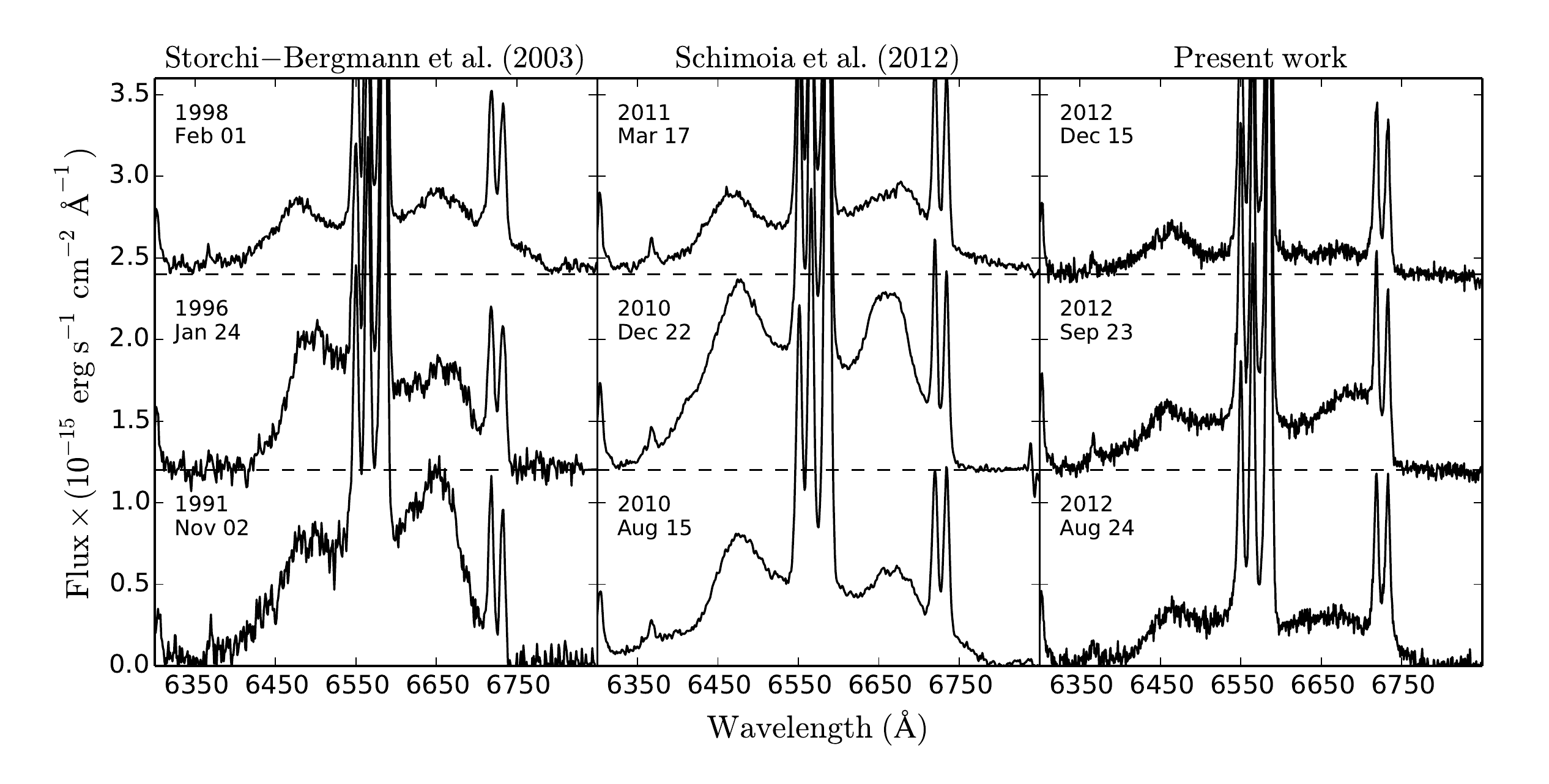}
\caption{ \emph{Left}:
Comparison of the relation between the integrated H$\alpha$ flux,
$F_{\rm broad}$, and velocity separation between the blue and red
peaks, $\Delta V = V_B - V_R$, for three different monitoring
campaigns.  The data from the present work are shown as blue circles;
the data from Storchi-Bergmann et al. (2003) are shown as red
triangles, and the data from Schimoia et al. (2012) are shown as green
squares.  The comparison shows that during this campaign
$F_{\rm broad}$ remained at the lowest fluxes ever seen while $V_B -
V_R$ was at a very large separation. This trend is in agreement with
that of the previous observations. The blue dotted line represents the linear regression
for the data from \cite{Schimoia} + that of this work (\mr{r\sim-0.343}); while the dot-dashed line
is the linear regression for the data from \citet{SB03} alone (\mr{r\sim-0.823}). The 
black dashed line is the linear regression for all the data together (\mr{r\sim-0.354}).
\emph{Right}: three representative profiles of each of the three time periods covered by our previous and present work.}

\label{fb_delv}
\end{figure*}

Figure \ref{fb_delv}
shows, in the left panel, a comparison of the measurements of $F_{\rm broad}$ and $\Delta V$ of 
the present and previous works. The right panel shows 3 typical profiles for each of the three epochs we have collected the data.

The left panel of Figure \ref{fb_delv} shows that in the current campaign,
the broad H$\alpha$ flux  $F_{\rm broad}$ reached historically low levels and remained low
throughout the campaign, with variations comparable to or just barely
larger than the uncertainties in the measurements.  The separation
$\Delta V$ between the blue and red peaks was also at the largest
values we have ever observed in this source, what is still consistent with the reverberation scenario.

We now investigate the significance of the inverse correlation
between the integrated flux of the 
H$\alpha$ double-peaked line and the velocity separation of the blue and red peaks. Fig. 5 (left) shows that there 
is a distinct behavior between the data from \citet{SB03} (red triangles) -- taken between 1991 and 2002, 
and that from our more recent studies, corresponding to the period between 2010 - 2013, 
including both the data from \citet{Schimoia} and that of the present work (blue circles and green squares).

Between these two data sets there is a temporal gap of approximately 7-8 years.
The right panels of Fig. 5 show that in the first data set  (1991--2002) the profile showed large variations, 
both in flux and peak separation, with a strong inverse correlation between the two.
In the more recent observations that begun in  2010, the profiles became broader.
In order to investigate the significance of the inverse correlation, we have grouped the data as 
follows: (i) only the data from \citet{SB03}; (ii) only the most recent data of  \citet{Schimoia} + 
that of the present work;  and (iii) all the data together. The resulting linear regressions are shown 
as dashed lines in Fig. \ref{fb_delv}.

We have calculated the correlation coefficient and respective significance for the three different groups of data above:
\begin{enumerate}
\item For the data from \citet{SB03} alone we found a linear correlation coefficient \mr{r=-0.823^{+0.034}_{-0.038}}, which implies significance levels from \mr{99.95\%} to \mr{99.99\%};

\item For the data from \citet{Schimoia} plus that of this work we found a correlation coefficient \mr{r=-0.343^{+0.049}_{-0.050}}, which 
gives significance levels from \mr{89.25\%} to \mr{97.17\%};

\item For all data together we found a correlation coefficient of \mr{r=-0.354^{+0.029}_{-0.029}}, and the respective significance levels
are \mr{97.05\%} and \mr{99.06\%}.
\end{enumerate}

Thus we conclude that for any set of data there is an inverse correlation with a significance
level \mr{\gtrsim 90 \%}, while for the 
complete data set the correlation is \mr{\sim 97\%}. The data are thus
consistent with a low-flux state corresponding to the widest
separation between the two peaks, which leads us to the conclusion
that the overall data are consistent with the reverberation scenario.

\subsection{An Accretion Disk Model}
\label{accretion_disk} In order to test further the scenario we have
proposed for the accretion disk \citep{Schimoia}, we modeled the
H$\alpha$ profiles using the accretion disk model described by
\citet{Gilbert99}, \citet{SB03}, and \citet{Schimoia}.  In this
formulation, the broad double-beaked emission line originates in a
relativistic Keplerian disk of gas surounding the SMBH. The
line-emitting portion of the disk is circular and located between an
inner radius $\xi_{1}$ and an outer radius $\xi_{2}$ (where $\xi$ is
the disk radius in units of the gravitational radius $r_g={GM_{\rm
BH}}/{c^2}$, $c$ is the speed of light, $G$ is the gravitational constant,
and $M_{\rm BH}$ is the mass of the black hole). The disk has an
inclination $i$ relative to the line of sight (i.e., zero degrees is
face-on). Superimposed on the axisymmetric emissivity of the circular
disk, there is a perturbation in the form of a spiral arm.  We adopted
the ``saturated spiral model'' \citep{Schimoia} for the total
emissivity of the accretion disk, as it best reproduces the data.
This emissivity law is given by
\begin{multline} \label{arm} \epsilon(\xi,\phi) = \epsilon (\xi) \left
\{ 1 + \frac{A}{2} \exp \left [ -\frac{4 \ln2}{\delta^{2}} (\phi -
\psi_{0})^{2} \right ] \right . \\ \left . + \frac{A}{2} \exp \left [
-\frac{4 \ln2}{\delta^{2}} (2\pi - \phi + \psi_{0})^{2} \right ]
\right \},
\end{multline} where
\begin{equation}
 \epsilon(\xi) =\left\{ \begin{array}{ll} \epsilon_{0}\xi^{-q_{1}} &
 ,\,\xi_{1} < \xi < \xi_{q}\\
 \epsilon_{0}{\xi_{q}}^{-(q_{1}-q_{2})}\xi^{-q_{2}} &,\,\xi_{q} < \xi
 < \xi_{2} \end{array} \right.
\end{equation} is the axisymmetric emissivity of the disk. The
parameter $\xi_{q}$ is the radius of maximum emissivity, or saturation
radius, at which the emissivity law changes; $q_{1}$ is the index of
the emissivity law for $\xi_{1} < \xi < \xi_{q}$; $q_{2}$ is the index
for $\xi_{q} < \xi < \xi_{2}$.
$A$ is the brightness contrast between the spiral arm and the
underlying disk, and the expression between square brackets represents
the decay of the emissivity of the arm as a function of the azimuthal
distance $\phi - \psi_{0}$ from the ridge line to both sides of the
arm, assumed to be a Gaussian function with ${\rm FWHM} = \delta$
(azimuthal width).

The relation between the azimuthal angle $\phi_{0}$ and the angular
position $\psi_{0}$ of the ridge of emissivity on the spiral arm is
given by
\begin{equation}
 \psi_{0} = \phi_{0} + \frac{\ln(\xi/\xi_{sp})}{\tan p},
\end{equation} where $\phi_0$ is the azimuthal angle of the spiral
pattern, $p$ is the pitch angle, and $\xi_{sp}$ is the innermost radius
of the spiral arm.

The specific intensity from each location in the disk, in the frame of
the emitting particle is calculated as
 \begin{equation}
  I(\xi,\phi,\nu_{e}) =
  \frac{\epsilon(\xi,\phi)}{4\pi}\frac{e^{-(\nu_{e}-\nu_{0})^{2}/{2\sigma^{2}}}}{(2\pi)^{1/2}\sigma},
 \end{equation} where $\nu_{e}$ is the emission frequency and
$\nu_{0}$ is the rest frequency corresponding to H$\alpha\ \lambda
6564.6$ and $\sigma$ is the local ``broadening parameter''
\citep{CeH89}.

We fixed the following model parameters: $q_{1}=-2.0$, $q_{2}=3.0$,
$\xi_{sp}=\xi_{1}$, $p=50^{\circ}$, $\delta=70^{\circ}$ and
$i=34^{\circ}$, which are the same values adopted in previous works
\citep{SB03, Schimoia}.  Thus, the only parameters we allowed to vary
are $A$, $\xi_q$ and $\phi_0$. In practice, $\xi_q$ regulates the
width of the profile: when $\xi_q$ is closer to $\xi_1$ the profile is
broader, when $\xi_q$ is closer to $\xi_2$ the profile becomes
narrower; $\phi_0$ is the orientation of the spiral pattern and by
allowing this parameter to increase monotonically, we can reproduce
the variation of the relative intensity of the blue and red peaks as
due to the rotation of the spiral arm.

\begin{figure*} \centering
\includegraphics[width=16.5cm]{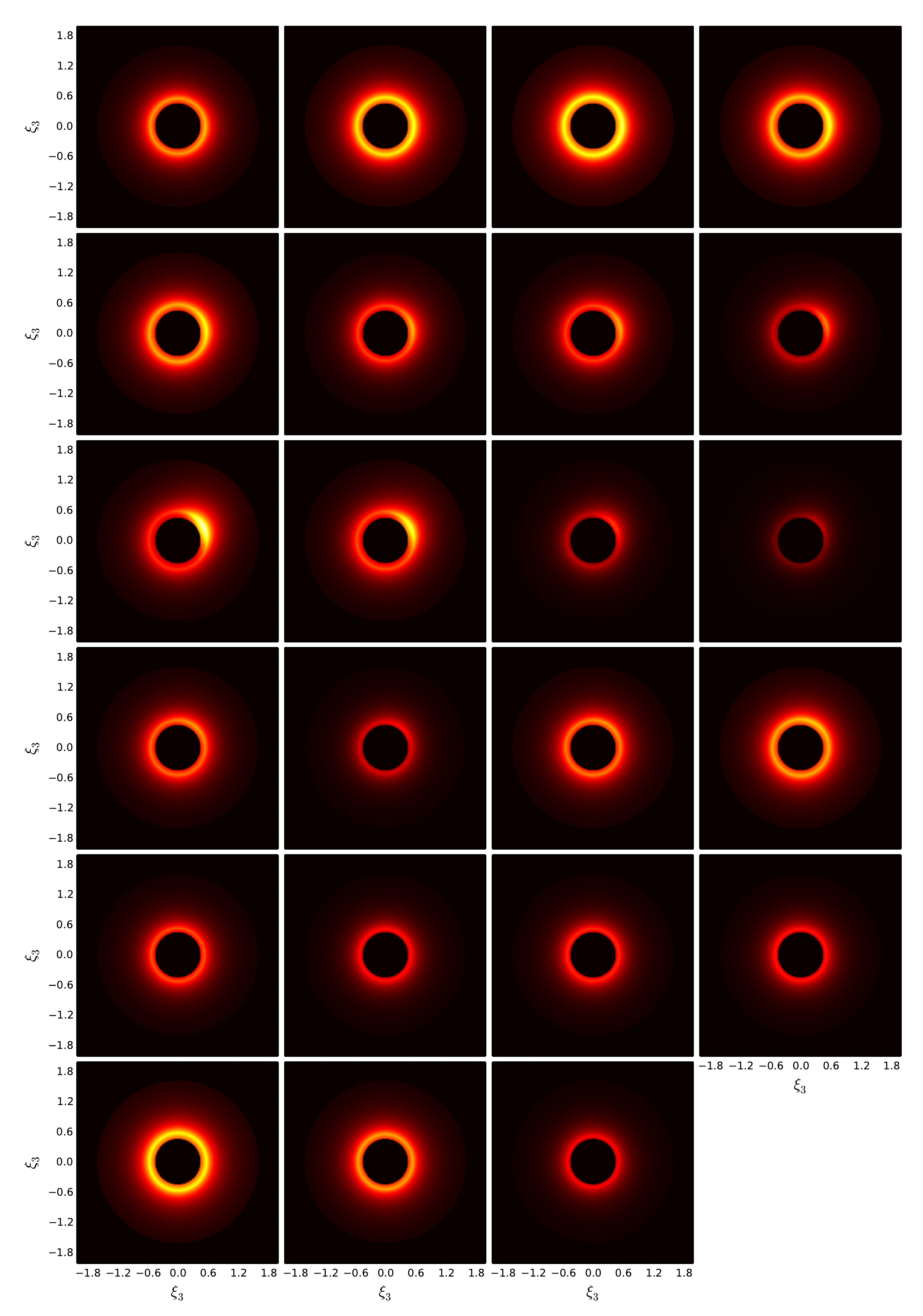}
\caption{ Image showing the disk emissivity utilized to model the
double-peaked profile for each epoch of observation of Figure
\ref{spectra}. White represents the brightest regions, and the
observer is to the bottom. The disk parameters for these epochs are
listed in Table \ref{pars}. On each frame the axis are expressed in
terms of $\xi_3=10^{-3}\xi$.}
\label{emissivity}
\end{figure*}

For direct comparison with the scenario proposed by \citet{Schimoia},
we adopt their derived angular velocity of the spiral arm, of
$\dot{\phi_0} \approx 0.680^{\circ}$\,day$^{-1}$, which corresponds to
a rotation period of $\sim 18$ months. Using this prescription, we
estimate the value of $\phi_0$ for the first optical observation
(MJD56147) as $775^{\circ}$, which means that the spiral arm should
have completed one rotation since the last observation of
\citet{Schimoia} for which $\phi_0 = 400^{\circ}$.  The best-fitting models
are compared with the observed line profiles in Figure \ref{spectra},
and the corresponding parameter values are listed in Table \ref{pars}.
The disk emissivity for each epoch is ilustrated in Figure
\ref{emissivity}.

\begin{deluxetable}{l c c c c c} \tablecolumns{5}
\tablecaption{Parameters of the Accretion Disk Model} \tablehead{ UT Date
&$\mathrm{MJD}$ &$\xi_q$ &$\phi_0 - 760^{\circ}$ &$A$\\
        & & &$(^{\circ})$ & }\\ 
        \startdata 
        2012 Aug 07 &56147.356 &550 &15 &0.2 \\
        2012 Aug 09 &56149.368 &570 &20 &0.2 \\ 
        2012 Aug 10 &56150.360 &580 &20 &0.2 \\ 
        2012 Aug 24 &56164.349 &580 &30 &0.5 \\
        2012 Aug 31 &56171.388 &570 &35 &0.5 \\ 
        2012 Sep 03 &56173.305 &550 &35 &0.8 \\ 
        2012 Sep 08 &56179.370 &550 &40 &0.8 \\ 
        2012 Sep 16 &56187.222 &540 &45 &2.0 \\ 
        2012 Sep 23 &56194.256 &580 &50 &2.5 \\
        2012 Sep 27 &56198.175 &570 &50 &1.5 \\ 
        2012 Oct 12 &56213.316 &525 &60 &1.5 \\ 
        2012 Oct 18 &56219.316 &500 &65 &1.5 \\ 
        2012 Oct 25 &56226.048 &545 &70 &0.3 \\ 
        2012 Oct 28 &56229.229 &515 &75 &0.5 \\
        2012 Nov 02 &56234.086 &545 &75 &0.2 \\ 
        2012 Nov 12 &56244.074 &560 &80 &0.2 \\ 
        2012 Nov 23 &56255.068 &530 &90 &0.0 \\ 
        2012 Dec 08 &56270.206 &515 &100 &0.0 \\ 
        2012 Dec 15 &56277.142 &520 &105 &0.0 \\
        2012 Dec 23 &56285.097 &515 &110 &0.0 \\ 
        2012 Dec 28 &56290.074 &570 &115 &0.0 \\ 
        2013 Jan 17 &56310.047 &545 &130 &0.0 \\ 
        2013 Jan 26 &56319.067 &510 &135 &0.0
\enddata
\label{pars}
\tablecomments{The date of observations are given in column (1) while column (2) gives the 
Modified Julian Date (JD$-2400000.5$). The resulting parameters of the modelling are given in the following columns: (3)
the radius of maximum emissivity of the accretion disk, (4) the azimutal orientation of the spiral arm and (5) the contrast 
between the spiral arm and the underlying disk.}
\end{deluxetable}

As can be seen in Figure \ref{spectra}, the model can indeed reproduce
the observed variations of the double-peaked profile.  We have tested
many values for $\xi_q$ in the range between the inner and outer
radius, $\xi_1=450$ and $\xi_2=1600$.  However, we achieved
the best fits by keeping $\xi_q$ close to the inner radius, in the
range $\xi_q\sim 500\,-\,600$.  This is consistent with the fact
that during the monitoring campaign we observed the broadest profiles
ever seen for this object.

In order to improve the fitting we also had to apply an overall blueshift 
of about $-250$\,km\,s$^{-1}$ to all model profiles.  Such a
velocity shift has been noted in our previous studies \citep{SB03,
Schimoia}, in which they were attributed to possible winds from the
accretion disk.  Another possible explanation for the velocity shifts
is the presence of the spiral pattern, since the spiral arm also
slightly modifies the velocity field and makes it depart from a pure
Keplerian velocity field.  We note that some epochs require blueshifts
somewhat higher or lower than $-250$\,km\,s$^{-1}$; however, the
corresponding improvements in the fit are smaller than the
uncertainties. Thus, we adopt the mean blueshift value of
$-$250\,km\,s$^{-1}$.

Modeling the present data revealed that the parameter $\xi_q$ --- the
``break" radius, corresponding to the region with strongest emission
in the disk --- must be kept close to the inner radius in order to
reproduce the observed double-peaked profiles. This result is
consistent with the scenario in which the central ionizing source is
faint and able to power just the innermost region of the accretion
disk.  The mean value of $\xi_q$ over all the observations is
$\overline{\xi_q} = 545 \pm 25$.  Assuming the black hole mass is $M_{\rm BH}
= 1.2\times10^8\,M_{\odot}$ \citep{Lewis06}, at this radius the
dynamical timescale of the accretion disk is $\tau_{dyn}\approx3$
months and the rotation period ($2\pi\tau_{dyn}$) is $\sim 18$ months,
which is consistent with the estimated rotation period for the spiral
arm. A property of the double-peaked profile that varies on this
timescale is the relative intensity of the blue and red peaks: for
instance, from MJD56194 to MJD56277 (see Figure \ref{spectra}) the
profile changed from a dominant red peak to a dominant blue peak in
$\sim 3$ months. We note that the light travel timescale,
at $\xi_q = 545$ is just 4 days, which is consistent with the upper
limit of $\sim 5$ days for the timescale for the H$\alpha$ flux
variations.

\subsection{Correlation Between X-Rays and H$\alpha$}
\label{xray}

\begin{figure} \centering
\includegraphics[width=8cm]{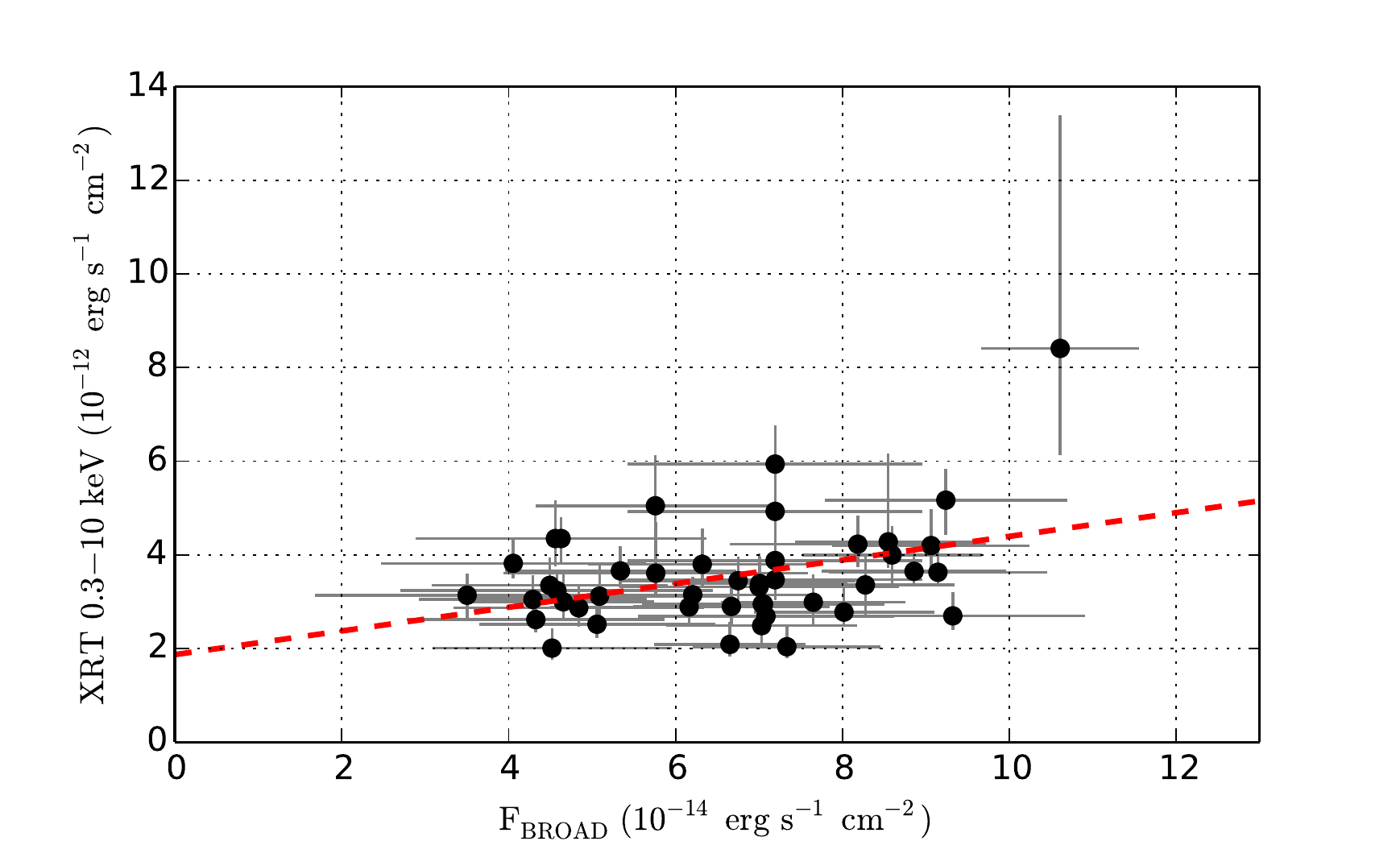}
\caption{correlation between X-ray and optical data, where black points
represent the fluxes of the X-ray band and the broad profile for the
same MJD (after interpolation of the optical light curve to match the
epochs of Swift observations).  The red dashed line is the best linear
correlation between the data. The median value of the correlation coefficient
is $\mathrm{r=0.253^{+0.105}_{-0.133}}$ which corresponds to a level of significance
of 56.8\% -- 98.4\%.}
\label{xray_opt}
\end{figure}

In order to look for a direct reverberation signal between the X-rays
(presumably emitted by the RIAF) and the response of the line-emitting
portion of the disk, we applied the interpolated cross-correlation
function method \citep[ICCF,][]{Gaskell, White, Peterson98,
Peterson04} between the X-ray and H$\alpha$ light curves.  
Unfortunately, we are unable to constrain the lag between
the X-ray and H$\alpha$ light curves. We are similarly unable to
constrain any possible lag between the X-rays and the UV continuum.
The main reason that the time-series analysis fails to yield results
is primarily the absence of strong flux variations in every band in
this campaign. It also seems apparent that,
at least in this low-luminosity state, the time sampling was not frequent enough.
This conclusion is supported by the fact that we could detect significant
variations in the X-rays and H$\alpha$ light curves on time intervals
as short as our mean sampling time interval.

We also investigated the degree to which the X-rays and 
H$\alpha$\ light curves are correlated at zero time delay. As the
H$\alpha$\ light curve is more sparsely sampled than the {\em Swift}/XRT
light curve, we carried out a piece wise linear interpolation of the H$\alpha$\ light
curve to match each of the {\em Swift} observations.  We were then
able to perform simple linear regression analysis of the
X-ray/H$\alpha$\ pairs.



In order to take into account the effect of the uncertainties of the fluxes measurements
in the correlation, we proceeded as follows: before calculating the correlation coefficient,
we scattered each flux measurement using a Gaussian probability distribution 
in which the mean value of the Gaussian corresponded to the value of the flux measurement
and the root mean square of the Gaussian corresponded to
mean squared value of the lower and upper uncertainties.
We repeated this process 10$^5$ times
and calculated the median value for the distribution
of values of the correlation coefficient as $\mathrm{\overline{r}=0.253^{+0.105}_{-0.133}}$.
The upper and lower limits correspond to $\pm$34.1\% of the 
distribution around the median. The lower and upper limits for the correlation coefficient
imply a significance level of 56.8\% to 98.4\%.

We can thus conclude that there is a marginal correlation between the  X-ray and H$\alpha$ flux variations; the variations are not completely random. Nevertheless the correlation is not tight enough to confirm that the two fluxes vary together. 
The weakness of the correlation is almost certaintly attributable to the  low amplitude of the variations
during the monitoring campaign. The lack of variability in the UV precluded any similar analysis in that band.

\subsection{Comparison with other objects and accretion disk models}

Monitoring of ``double-peaked'' emitters have been used as a probe 
for testing accretion disk models and to understand what are the physical mechanisms
responsible for the observed changes in the double-peaked profiles.
In this subsection we make a comparison between the observed properties of 
the NGC1097 double-peaked profile and two other also very well studied 
double-peaked emitters: Arp\,102B and 3C390.3.

In the case of 3C\,390.3, many works have reported spectroscopic long-term monitoring (years do decades) \citep{Shapovalova01,Sergeev02, Shapovalova10} 
or short-term monitoring (a few months) \citep{Dietrich} of the AGN optical continuum around 5100\,\AA\ and the variations of the 
broad Balmer double-peaked emission lines, H$\alpha$ and H$\beta$. Common to these works is the conclusion that the
broad double-peaked Balmer lines flux variations do follow the optical continuum variations with time delays of
 $\tau$(H$\alpha$)$\gtrsim$\,56 days and $\tau$(H$\beta$)$\gtrsim$\,44 days. Also, \cite{Dietrich98}
 found that the flux variations of the Balmer lines of Arp 102B seems to be correlated with
 the X-ray flux variations with a time delay of $\sim$\,20 days. 
However, for the case Arp\,102B the variations of the flux of the Balmer broad double-peaked emission lines do not appear
to be strongly correlated with the variations of the AGN optical continuum \citep{Shapovalova13}. In spite of the
weak correlation, the authors estimate a time delay of $\tau(H\beta)\sim$ 20 days.

In the case of NGC\,1097, we were never able to detect the AGN optical continuum \citep{SB93, SB03, Schimoia}, 
and this was the motivation for us to look for the continuum in X-rays and UV. Unfortunately, 
the variations were small during our campaign, but in \S\ref{xray} we have nevertheless shown that  
the profile variations are weakly correlated with the X-ray continuum variations, indicating that 
the X-rays could indeed be the driver of the double-peaked profile.

In the case of Arp 102B the variations of the relative intensity of 
the blue and red peaks were first modeled by \citet{Newman97} in the context of a relativistic accretion disk
in which the non axisymmetric emissivity part of the disk has the shape of a \emph{hot spot} that
rotates in the disk around the central SMBH in the dynamical timescale. In the case of 3C390.3 the \emph{hot spot}
model have been used to explain the observed flux ``outbursts'' of the blue peak of the profile \citep{Jova10}.
\citet{Lewis10} have reported a long-term monitoring of a few radio-galaxies
in which the double-peaked profiles display small flux {\it bumps} that
move in the velocity space and are attributed to hot spots in the accretion disk.
Nevertheless, in almost all cases the hot spots are short lived: they have been observed to appear and are more likely
to last from several months to a few years. In contrast, the alternating relative intensity of the blue and red
peaks of NGC1097 have lasted for more than 20 years,
suggesting that the features producing the variation of the relative intensity 
of the blue and red peaks are longer lived than the hot spots observed in other double peaked emitters. 
The variations in the relative intensity of the blue and red peaks in NGC\,1097 have been clearly 
observed both when the flux of the
broad line is in a high state (higher line flux and narrower profile)
as well as when the flux is in a low state (low line flux and broader profile). 
On the other hand, we note that the emissivity of the \emph{saturated spiral model} adopted in our 
modelling is not that different from that of a \emph{hot spot}, in the sense that the spiral arm + 
broken power law emissivity component can effectively concentrate the non-axisymmetric emissivity 
in a structure very similar to a hot spot (see Fig. \ref{emissivity}, MJD56194).
Nevertheless, we favor the scenario of the \emph{saturated spiral model} for NGC1097
due the fact that the fast timescale variations of the profile, namely in the 
the velocity separarion of the peaks and integrated flux of the broad profile, occur in
timescales of a few days to a few weeks. At these short time scales, these variation can only be attributed to changes in 
the illumination/ionization of the disk, which have been modeled by changing the break radius, $\xi_q$, while the
the rotation of the non axissymetric pattern allow us to simultaneously model the variations of the relative intensity of the
blue and red peaks.

Finally we point out that at some epochs there is a flux excess in the high velocities of the blue wing of the double-peaked profile 
(see Fig. \ref{spectra}, for instance MJD56194 and 56290), that was also observed in previous works \citep{SB03,Schimoia}.
The origin of this excess is not completely understood. In \cite{SB03} it was proposed 
that it could be due to the presence of a wind emanating from the disk, what is not taken into account in the model.

\section{Conclusions}
We monitored the AGN in NGC\,1097 for $\sim$6 months, from 2012 August to 2013 February,
in three spectral bands: in the X-ray (0.3 -- 10 keV), using the {\em Swift}/XRT, in the UV, using the {\em Swift}/UVOT telescope, and in the  double-peaked H$\alpha$ profile, with data from the SOAR and Gemini-South telescopes. Our main conclusions are:

\begin{enumerate}
\item The X-ray fluxes varied on the shortest timescale probed by the
  observations, 4 days, indicating that the emitting structure has a
  size smaller than 4 light days; the maximum variation observed between two consecutive dates was
  97\%, but the average variation over the whole period was only
  $13\%$.

\item The H$\alpha$ flux also varied on the shortest timescale probed
  by the observations, $\sim 5$ days. The maximum variation observed
  was 90\%, but the average variation over the whole period was 20\%.

\item To within the uncertainties, the UV flux did not vary. This can
  be due to the contribution of the compact nuclear starburst
  previously found within the inner 9\,pc, that dominates the UV
  continuum emission, and is probably diluting the contribution of a
  variable source.

\item Although we have not find a clear reverberation signal 
  in our most recent data, when we consider this data together 
  with the previous ones from epochs 1991-2001 and 2010-2012, 
  they are consistent with the reverberation scenario, as we 
  can still observe, in the overall data, an inverse correlation 
  between the H$\alpha$ flux and the separation between the 
  red and blue peaks of the H$\alpha$ profile, as noted in 
  previous studies.

\item We were able to reproduce the variations of the double-peaked
  profile using the same accretion disk model of \citet{Schimoia}; in
  particular, we kept the same inner and outer radii, as well as the
  18-month rotation period for a spiral arm in the disk, that,
  combined with variations in the contrast of the spiral arm,
  successfully reproduces the variation of the relative intensity of
  the blue and red peaks.

\item The small range of variation of the width of the double-peaked
  profiles was reproduced by keeping the $\xi_q$ values (radius of
  maximum emissivity) close to the inner radius. This radius corresponds
  to a rotation period of $18$ months and a light travel time of
  $\sim 4$ days, in agreement with the shortest variation
  timescales observed.

\item We find only a marginal correlation between the X-ray and
  H$\alpha$ flux variations, although the analysis
  reveals the need for more frequent monitoring of this source,
  particularly when it is in such a low-luminosity state.

\end{enumerate}

The presence of some correlation between the X-ray flux and the broad 
H$\alpha$ profile, even though we could get only an upper limit on the 
shortest variation time scale, leads us to the conclusion that the 
variations of the broad H$\alpha$ line profile do follow the variations 
of the X-ray flux. But we also note that the H$\alpha$ flux remained at the
lowest levels we have ever observed, and the blue and red peaks of the
double-peaked profile remaining at the widest separation. This
indicates that we have caught the AGN in NGC\,1097 in a very low
activity state, in which the ionizing source was very weak and capable
of ionizing just the innermost part of the gas in the
disk. Nonetheless, the data presented here still support the picture
in which the gas that emits the double-peaked Balmer lines is
illuminated/ionized by a source of high-energy photons which is
located interior to the inner radius of the line-emitting part of the disk.

%
%

\acknowledgments
J.S. acknowledges CNPq, National Council for Scientific and Technological Development - Brazil,
for partial financial support and The Ohio State University by the hospitality.
This research has made use of the
  XRT Data Analysis Software (XRTDAS) developed under the responsibility
  of the ASI Science Data Center (ASDC), Italy.
  This research has made use of data obtained through the High Energy 
Astrophysics Science Archive Research Center Online Service, provided by the 
NASA/Goddard Space Flight Center.
At Penn State D.G. and M.E. acknowledge support from the NASA Swift program
through contract NAS5-00136. Additionally, M.E. acknowledges the warm hospitality of the Center for Relativistic Astrophysics 
at Georgia Tech and the Department of Astronomy at the University of Washington. B.M.P. is grateful for support by the NSF
through grant AST-1008882 to The Ohio State University.


\end{document}